\documentclass{aa}
\usepackage{natbib,twoopt}
\usepackage[utf8]{inputenc}
\usepackage{graphicx}
\usepackage{subcaption}
\usepackage{amsmath}
\usepackage{amssymb}
\usepackage{multirow}
\usepackage{siunitx}
\usepackage[breaklinks=true]{hyperref} 
\hypersetup{
  colorlinks=true,   
  urlcolor=cyan,     
  linkcolor=blue,     
  citecolor=cyan
}
\usepackage[varg]{txfonts}
\bibpunct{(}{)}{;}{a}{}{,} 
\makeatletter
\newcommandtwoopt{\citeads}[3][][]{\href{http://adsabs.harvard.edu/abs/#3}%
{\def\hyper@linkstart##1##2{}%
\let\hyper@linkend\@empty\citealp[#1][#2]{#3}}}
\newcommandtwoopt{\citepads}[3][][]{\href{http://adsabs.harvard.edu/abs/#3}%
{\def\hyper@linkstart##1##2{}%
\let\hyper@linkend\@empty\citep[#1][#2]{#3}}}
\newcommandtwoopt{\citetads}[3][][]{\href{http://adsabs.harvard.edu/abs/#3}%
{\def\hyper@linkstart##1##2{}%
\let\hyper@linkend\@empty\citet[#1][#2]{#3}}}
\newcommandtwoopt{\citeyearads}[3][][]%
{\href{http://adsabs.harvard.edu/abs/#3}
{\def\hyper@linkstart##1##2{}%
\let\hyper@linkend\@empty\citeyear[#1][#2]{#3}}}
\makeatother

\def\Fe II{\ion{Fe}{II}}

\begin{document}

\title{Connecting X-ray nuclear winds with galaxy-scale ionised outflows in two $z\sim1.5$ lensed quasars}

\author{G.~Tozzi\inst{\ref{inst1}, \ref{inst2}}\and G.~Cresci\inst{\ref{inst2}}\and A.~Marasco\inst{\ref{inst2}}\and E.~Nardini\inst{\ref{inst1}, \ref{inst2}}\and A.~Marconi\inst{\ref{inst1}, \ref{inst2}}\and F.~Mannucci\inst{\ref{inst2}}\and G.~Chartas\inst{\ref{inst3}}\and F.~Rizzo\inst{\ref{inst4}}\and A.~Amiri\inst{\ref{inst1}, \ref{inst2}}\and M.~Brusa\inst{\ref{inst5}, \ref{inst6}}\and A.~Comastri\inst{\ref{inst6}}\and M.~Dadina\inst{\ref{inst6}}\and G.~Lanzuisi\inst{\ref{inst6}}\and V.~Mainieri\inst{\ref{inst7}}\and M.~Mingozzi\inst{\ref{inst8}}\and M.~Perna\inst{\ref{inst9}, \ref{inst2}}\and G.~Venturi\inst{\ref{inst10}, \ref{inst2}}\and C.~Vignali\inst{\ref{inst5}, \ref{inst6}}}

\institute{Dipartimento di Fisica e Astronomia, Università di Firenze, Via G. Sansone 1, 50019, Sesto Fiorentino (Firenze), Italy\\ \email{giulia.tozzi@unifi.it}\label{inst1} \and INAF - Osservatorio Astrofisico di Arcetri, Largo E. Fermi 5, 50127, Firenze, Italy\label{inst2} \and Department of Physics and Astronomy, College of Charleston, Charleston, SC, 29424, USA\label{inst3} \and Max-Planck Institute for Astrophysics, Karl-Schwarzschild Str. 1, D-85748, Garching, Germany\label{inst4} \and Dipartimento di Fisica e Astronomia dell’Università degli Studi di Bologna, via P. Gobetti 93/2, 40129 Bologna, Italy\label{inst5} \and INAF - Osservatorio di Astrofisica e Scienza dello Spazio, via P. Gobetti 93/3, 40129, Bologna, Italy\label{inst6} \and European Southern Observatory, Karl-Schwarzschild-Str. 2, D-85748 Garching, Germany\label{inst7} \and Space Telescope Science Institute, 3700 San Martin Drive, Baltimore, MD 21218, USA\label{inst8} \and Centro de Astrobiología, (CAB, CSIC-INTA), Departamento de Astrofísica, Cra. de Ajalvir Km. 4, 28850 – Torrejón de Ardoz,
Madrid, Spain\label{inst9} \and Instituto de Astrofísica, Facultad de Física, Pontificia Universidad Católica de Chile, Casilla 306, Santiago 22, Chile\label{inst10}
}
\abstract {} {Outflows driven by active galactic nuclei (AGN) are expected to have a significant impact on host galaxy evolution, but the matter of how they are accelerated and propagated on galaxy-wide scales is still under debate. This work addresses these questions by studying the link between X-ray, nuclear ultra-fast outflows (UFOs), and extended ionised outflows, for the first time, in two quasars close to the peak of AGN activity ($z\sim2$), where AGN feedback is expected to be more effective.} {Our selected targets, HS 0810+2554 and SDSS J1353+1138, are two multiple-lensed quasars at $z\sim1.5$ with UFO detection that have been observed with the near-IR integral field spectrometer SINFONI at the VLT. We performed a kinematical analysis of the [O III]$\lambda$5007 optical emission line to trace the presence of ionised outflows.} {We detected spatially resolved ionised outflows in both galaxies, extended more than 8 kpc and moving up to $v>2000$ km/s. We derived mass outflow rates of $\sim$12 \(\text{M}_\odot\)yr$^{-1}$ and $\sim$2 \(\text{M}_\odot\)yr$^{-1}$ for \object{HS 0810+2554} and \object{SDSS J1353+1138}.} {Compared with the co-hosted UFO energetics, the ionised outflow energetics in HS 0810+2554 is broadly consistent with a momentum-driven regime of wind propagation, whereas in SDSS J1353+1138, it differs by about two orders of magnitude from theoretical predictions, requiring either a massive molecular outflow or a high variability of the AGN activity to account for such a discrepancy. By additionally considering our results together with those from the small sample of well-studied objects (all local but one) having both UFO and extended (ionised, atomic, or molecular) outflow detections, we found that in 10 out of 12 galaxies, the large-scale outflow energetics is consistent with the theoretical predictions of either a momentum- or an energy-driven scenario of wind propagation. This suggests that such models explain the acceleration mechanism of AGN-driven winds on large scales relatively well.}
\maketitle
\section{Introduction}     \label{sec:introduction}
Feedback mechanisms from active galactic nuclei (AGN) are widely considered to play a key role in galaxy formation and evolution (\citeads{1998A&A...331L...1S, 2010MNRAS.402.1516K, 2012ARA&A..50..455F, 2015ARA&A..53..115K}). AGN feedback is indeed included in all theoretical, semianalytic, and numerical studies of galaxy formation and evolution (e.g. \citeads{2004ApJ...600..580G, 2005Natur.433..604D, 2010ApJ...717..708C, 2016MNRAS.461.3457G, 2019MNRAS.490.3234N}) as it allows us to reconcile theoretical predictions with observed galaxy properties.
Such AGN activity is considered the main factor responsible for the quenching of star formation in more massive galaxies  (so-called ‘negative feedback’). The energy output of a supermassive black hole (SMBH) accreting close to the Eddington limit is large enough to drive massive, wide-angle outflows on large scales (e.g. \citeads{2012ApJ...745L..34Z, 2014MNRAS.439..400Z, 2014MNRAS.444.2355C, 2015ARA&A..53..115K, 2017MNRAS.465..547P}) that are capable of either sweeping the gas out of the host galaxy, thus reducing the host galaxy gas reservoir for star formation, or heating the intergalactic medium of the host galaxy through the injection of thermal energy, thus preventing the gas from cooling and collapsing to form stars (‘ejective’ versus ‘preventive’ feedback; e.g. see \citeads{2017ApJ...839..120W, 2018NatAs...2..179C}). Both processes are expected to halt the accretion onto the central black hole and, consequently, to give rise to the SMBH mass values observed to correlate with the physical properties of the host galaxy bulge (i.e. its mass, velocity dispersion and luminosity; e.g. \citeads{2000ApJ...539L...9F, 2013ARA&A..51..511K}). Additional observational evidence supporting the mutual influence of the central SMBH and the host galaxy comes from the observed similarity between the star formation (SF) and BH accretion histories across cosmic time. In fact, both activity histories are seen to peak at $z\sim2$ \citepads{1996MNRAS.283.1388M, 2014ARA&A..52..415M}, meaning that the bulk of both SF and BH accretion occurred within $z\sim1-3$ (e.g. \citeads{2004MNRAS.351..169M, 2010MNRAS.401.2531A, 2015MNRAS.451.1892A}). The epoch $z\sim1-3$ (also referred to as ‘cosmic noon’) is hence crucial to the study of such phenomena and their effects in action since this is the time when AGN feedback is expected to be more effective.

Thanks to advanced integral-field spectroscopic (IFS) facilities, AGN-driven outflows have been exhaustively observed from optical to IR and mm bands, in both local (e.g. \citeads{2010A&A...518L.155F, 2015A&A...583A..99F, 2012A&A...543A..99C, 2014A&A...562A..21C}) and high-redshift galaxies (e.g. \citeads{2012A&A...537L...8C, 2012MNRAS.425L..66M, 2015A&A...580A.102C, 2015ApJ...799...82C}). It is worth mentioning that while theoretical predictions usually refer to the whole outflowing gas, observations usually trace the emission produced by a single gas phase of the outflow. Therefore, in order to properly compare model predictions with observational results, it is fundamental to obtain a complete, multi-phase description of the outflow (e.g. \citeads{2018NatAs...2..176C, 2018NatAs...2..198H}).

Even though the existence of AGN-driven outflows has been widely confirmed thanks to observations, there are a number of relevant open questions that remain unanswered, considering, for instance, how the energy released by the accreting BH is coupled with the interstellar medium (ISM), thus driving large-scale outflows, and how efficient the coupling is between the nuclear and galaxy-scale outflows.

Theoretical models (e.g. \citeads{2003ApJ...596L..27K, 2005ApJ...635L.121K, 2015ARA&A..53..115K}) predict fast ($v\sim0.1c$), highly ionised winds, accelerated on sub-pc scales by the AGN radiative force as the origin of the strong galaxy-scale feedback. As the nuclear wind impacts the ISM of the host galaxy, it produces an inner reverse shock that slows down the wind, along with an outer forward shock accelerating the galactic ISM. Depending on the efficiency of cooling processes (typically radiative) in removing energy from the hot shocked inner gas, there are two main wind driving-modes (e.g. \citeads{2010MNRAS.408L..95K, 2014MNRAS.444.2355C, 2015ARA&A..53..115K}). If the cooling occurs on a timescale that is shorter than the wind flow time, most of the inner wind kinetic energy is lost (usually via inverse Compton scattering) and, therefore, the wind momentum is the only conserved physical quantity (‘momentum-driven’ regime). Vice versa, if the cooling is negligible, the postshock gas retains all the mechanical energy and expands adiabatically (‘energy-driven’ regime), sweeping up a significant amount of the host galaxy gas. According to a widely accepted picture, the observed scaling relations are the result of the effect of AGN feedback acting on the host galaxy through two distinct, subsequent phases (e.g. \citeads{2012ApJ...745L..34Z, 2015ARA&A..53..115K}): an initial momentum-driven regime lasting so long as the BH mass has not yet reached the $M_{\rm BH}-\sigma$ relation and the outflow is confined within $\sim$1 kpc from the central BH, followed by a later energy-driven phase after the BH mass has settled on the relation, during which the outflow can propagate beyond 1 kpc scales.

From an observational point of view, the most promising candidates for acting as the ‘engine’ of the large-scale feedback are the so-called ultra-fast outflows (UFOs), which are highly-ionised, accretion disk winds with mildly relativistic velocities that originate at sub-pc scales. They are usually detected in AGN X-ray spectra \citepads{2002ApJ...579..169C, 2016ApJ...824...53C, 2013MNRAS.430...60G, 2015MNRAS.451.4169G, 2015Sci...347..860N} via the presence of strongly blueshifted absorption lines of highly-ionised metals (typically iron, e.g. Fe XXV and Fe XXVI). UFOs are found in at least 40\% of the local sources \citepads{2010A&A...521A..57T, 2011ApJ...742...44T, 2012MNRAS.422L...1T, 2013MNRAS.430.1102T}, with typical mass outflow rates of $\sim0.01-1$ \(\text{M}_\odot\) yr$^{-1}$ and kinetic powers of log$\dot{E}_{\rm K}\sim42-45$ erg s$^{-1}$ \citepads{2012MNRAS.422L...1T}.

Moving to high redshifts ($z>1$), the UFO detection is hampered by the resolution or sensitivity limits of the current observational facilities. In fact, the number of AGN hosting UFOs at $z>1$ that have been discovered thus far drastically decreases to 14 objects (see \citeads{2018A&A...610L..13D} on the list of published objects, and \citealt{Chartas.high-z.UFOs} for the latest updates); of these, twelve are gravitationally lensed systems (\citeads{2002ApJ...573L..77H, 2003ApJ...595...85C, 2007ApJ...661..678C, 2009NewAR..53..128C, 2016ApJ...824...53C, 2018A&A...610L..13D}; \citealt{Chartas.high-z.UFOs}). Strong gravitational lensing is indeed a well-known, powerful tool to investigate the physical properties of distant quasars (QSOs). The magnified view delivered by gravitational lenses allows us to separate the active nuclei from their hosts, enabling new measurements and spatially resolved studies, which otherwise would not have been possible beyond the local Universe (e.g. \citeads{2006ApJ...649..616P, 2009ApJ...702..472R, 2017ApJ...845L..14B,2020MNRAS.495.2387S,2021MNRAS.500.3667S}).

To test theoretical predictions and to shed light on the acceleration and propagation mechanisms of large-scale outflows, we need to compare the energetics of the X-ray nuclear UFO with that of the large-scale outflow. In this work, we focus on the ionised phase of large-scale outflows traced by the optical emission of the [O III]$\lambda \lambda$4959,5007 line doublet. Given that it is a forbidden transition, it preferentially traces the emission originating from the kpc-scale, typical of the AGN Narrow Line Region (NLR), since it cannot be produced on the high-density ($n\sim10^{10}$ cm$^{-3}$), sub-pc scale of the AGN broad line region (BLR). In the presence of outflows, the [O III] line profile is highly asymmetric with a broad, blueshifted wing corresponding to high speeds along the line of sight ($v\gtrsim1000$ km s$^{-1}$ (see e.g.  \citeads{2015A&A...580A.102C, 2015ApJ...799...82C, 2015A&A...574A..82P, 2016A&A...588A..58B, 2020A&A...644A..15M}).

This work is  aimed at studying the connection between nuclear, X-ray UFOs, and the ionised phase of large-scale outflows, for the first time, in two QSOs close to the peak of AGN activity ($z\sim2$). We use the [O III]$\lambda$5007 emission line to trace the ionised outflow, along with results on X-ray UFOs from the literature to properly compare the wind energetics on different scales. In doing so, we aim to highlight (and take advantage of) the crucial role of gravitational lensing as powerful tool for overcoming the current observational limits and going on to investigate the physical properties of distant QSOs.
 
This paper is organised as follows. In Sect. \ref{sec:description}, we present the selected targets and describe the observation and data reduction procedure. In Sect. \ref{sec:analysis}, we show our data analysis and spectral fitting. The inferred results are then presented in Sect. \ref{sec:results}. In Sect. \ref{sec:discussion}, we discuss the wind acceleration mechanism in our two QSOs and compare our results with  findings from the literature. Finally, in Sect. \ref{sec:conclusion}, we outline our conclusions. We adopt a $\Lambda$CDM flat cosmology with $\Omega_{\rm m,0} = 0.27$, $\Omega_{\rm \Lambda,0} = 0.73$ and $H_{\rm 0} =70$ km s$^{-1}$ Mpc$^{-1}$  throughout this work.

\section{Description of the observed QSOs} \label{sec:description}
\subsection{Selection of targets} \label{sec:sample_sel}
Our sample consists of two $z\sim1.5$ multiple lensed QSOs, HS 0810+2554 and SDSS J1353+1138, observed with the Spectrograph for INtegral Field Observations in the Near Infrared (SINFONI, \citeads{2003SPIE.4841.1548E}) at the ESO Very Large Telescope (VLT) Unit Telescope 3 (UT3) within the framework of the program 0102.B-0377(A) (PI: G. Cresci). These objects were specifically selected as they are known to host UFOs (\citeads{2016ApJ...824...53C}; \citealt{Chartas.high-z.UFOs}) and to be at redshifts ($z\sim1.5$ and $z\sim1.6$ for HS 0810+2554 and SDSS J1353+1138, respectively), such that the optical rest-frame emission (tracing ionised outflows) falls in the range of wavelengths observed by SINFONI, namely, in the near-IR \text{J}-band ($\lambda \sim 1.1-1.4$ $\mu$m). Consequently, this selection in redshifts corresponds to study objects at epochs close to the peak of AGN activity ($z\sim2$).
In total, there are fourteen high redshift ($z > 1$) QSOs with UFO detection, of which seven are found in the literature (among them, HS 0810+2554; see \citeads{2018A&A...610L..13D} for an updated list), while the rest have not yet been published (including SDSS J1353+1138; \citealt{Chartas.high-z.UFOs}). These are among the brightest - in terms of 2--10 keV luminosity ($L_{2-10~\text{keV}}>10^{45}$ erg s$^{-1}$, except for PID352 $L_{2-10~\text{keV}}\sim10^{44}$ erg s$^{-1}$) - QSOs at high redshift; this is either because they are intrinsically luminous (of the total 14-QSO sample, only HS 1700+6416 and PID352; \citeads{2012A&A...544A...2L, 2015A&A...583A.141V}) or because they are subject to gravitational lens magnification (including APM 08279+5255, PG1115+080, H1413+117, HS 0810+2554 and MG J0414+0534\footnote{Here, we list only the sources with published results on UFO detection, but also the remaining unpublished objects are known to be gravitationally lensed \citep{Chartas.high-z.UFOs}.}; \citeads{2002ApJ...573L..77H, 2003ApJ...595...85C, 2007ApJ...661..678C, 2009NewAR..53..128C, 2016ApJ...824...53C, 2018A&A...610L..13D}). Thus, these objects deliver high quality X-ray spectra which clearly exhibit UFO absorption features, in spite of their high redshift ($z>1$). Amongst this $z>1$ sample, APM 08279+5255 ($z\sim3.9$; \citeads{2002ApJ...573L..77H}) has been the only $z>1$ QSO known to host a large-scale (molecular) outflow \citepads{2017A&A...608A..30F}, hence, this is where a connection between UFO and galaxy-scale outflow has been put forward thus far. Therefore, within the field of large-scale ionised outflows in HS 0810+2554 and in SDSS J1353+1138, this work will extend the number of $z>1$ QSOs for which the connection between nuclear and large scales has been accessed.
In the following, we provide a short description of HS 0810+2554 and SDSS J1353+1138, with their main properties listed in Table \ref{tab:sample}.
\begin{table*}[htp]
\centering
\begin{tabular}{c|cccc}
\hline
\hline
  Target name & $\alpha$(J2000) & $\delta$(J2000) & $z^{\rm a}$ & scale \\
 
 \hline
 HS 0810+2554 & $08^{\rm h}13^{\rm m}31^{\rm s}.3$ & $+25^{\circ}45'03''$ & $1.508\pm0.002$ & 8.67 kpc/$''$\\ 
 SDSS J1353+1138 & $13^{\rm h}53^{\rm m}06^{\rm s}.34$ & $+11^{\circ}38'04''.7$ & $1.632\pm0.002$       & 8.69 kpc/$''$\\ 
  \hline
\end{tabular}%
\caption{{\small Properties of our two-QSOs sample. $^{\rm a}$Redshifts are measured from the [O III] systemic component in integrated spectra extracted from the nuclear region of both sources (Sects. \ref{sec:BLR_fit} and \ref{sec:double_BLR}).}}
\label{tab:sample}
\end{table*}%
\subsubsection{HS 0810+2554}    \label{sec:HS}
\begin{figure}
   \resizebox{\hsize}{!}{\includegraphics{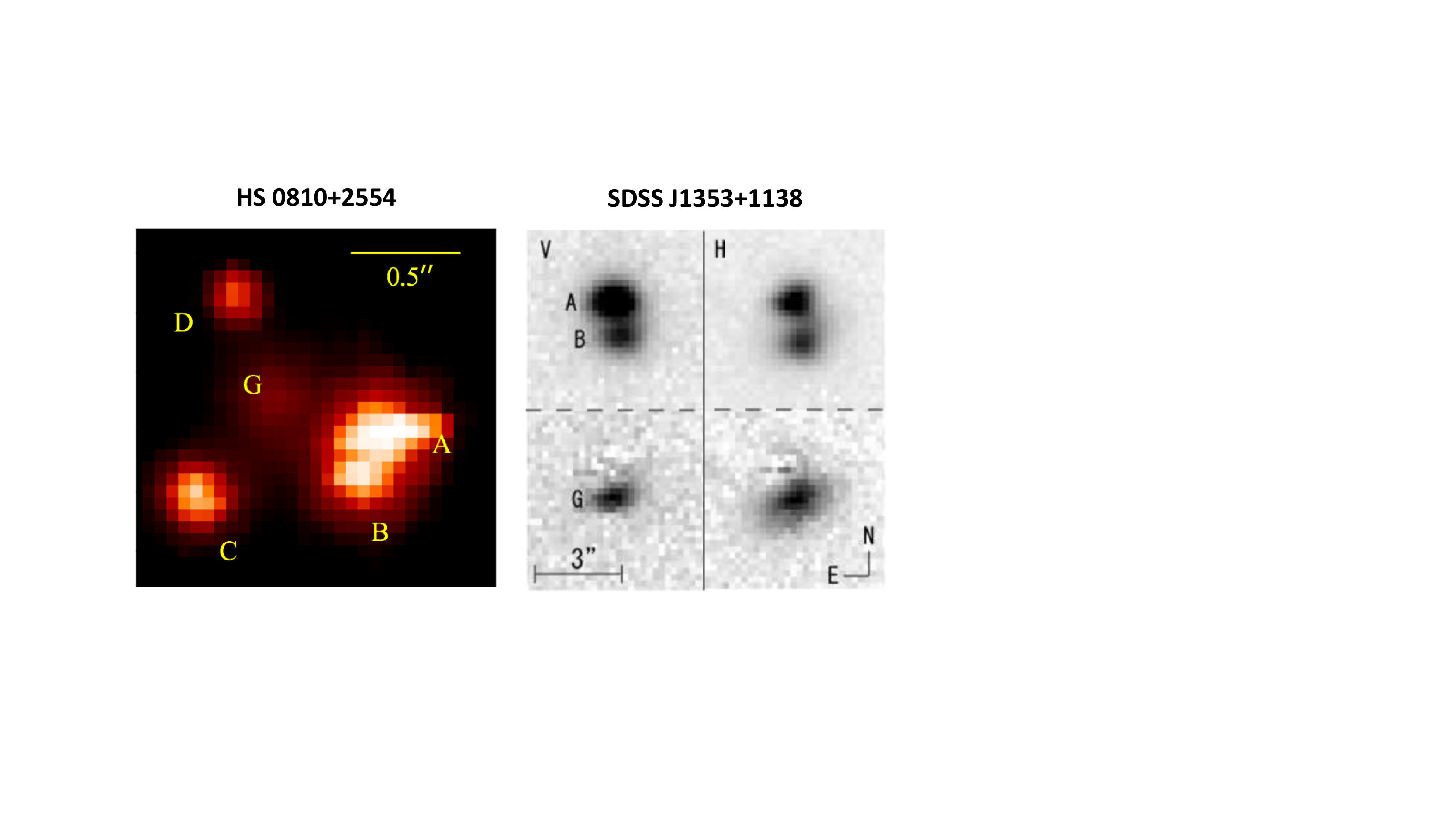}}
     \caption{{\small Lensed images of HS 0810+1154 (\textit{left}) and SDSS J1353+1138 (\textit{right}). \textit{Left:} \textit{HST} ACS F555W image of HS 0810+2554 showing the four magnified images of the background quasar in fold lens configuration: the C and D images are spatially resolved, while the pair A+B is blended together. At the centre, we can see the emission from the foreground lens galaxy. \textit{Right:} \textit{V} and \textit{H}-band images of SDSS J1353+1138 taken at the UH88 telescope (\textit{upper} panels) and corresponding images after the subtraction of A and B components (\textit{lower} panels), clearly showing  the lens galaxy (component G). Image from \citetads{2006ApJ...653L..97I}.}}
     \label{fig:opt_images}
\end{figure}

HS 0810+2554 is a radio-quiet, narrow absorption line (NAL; FWHM$~\lesssim500$ km s$^{-1}$) QSO at $z\sim1.5$, which was discovered by \citetads{2002A&A...382L..26R}. It consists of four lensed images in a typical fold lens configuration with the two southern, brightest images in a merging pair configuration (A+B), as shown in the \textit{HST} image in Fig. \ref{fig:opt_images} (\textit{left} panel). The lens galaxy (labelled with G) is detected in the \textit{HST} image, and its redshift is estimated to be $z_{\rm l}\sim0.89$ from the separation and the redshift distribution of existing lenses \citepads{2011ApJ...738...96M}. Quadruply lensed QSOs occur in strong gravitational lensing regimes (e.g. \citeads{1996astro.ph..6001N}), when the compact and bright UV accretion disk and X-ray corona emission regions overlap the lens caustics. This leads to high magnification factors, whose values strongly depend on the image and lens positions. As a consequence, a small change in the input parameters to the lens models (image and lens positions) can lead to a significant change in the image magnifications. For HS 0810+2554, estimates of the magnification factor $\mu$ in different spectral bands are found in the literature, in particular for the X-ray ($\mu \sim103$; \citeads{2016ApJ...824...53C}), optical ($\mu \sim120$; \citeads{2020MNRAS.492.5314N}), and radio emission ($\mu \sim25$; \citeads{2015MNRAS.454..287J}).

HS 0810+2554 was singled out as an exceptionally X-ray bright lensed object during an X-ray survey of NAL-AGN with outflows of UV absorbing material \citepads{2009NewAR..53..128C}. More recent \textit{Chandra} and \textit{XMM-Newton} observations \citepads{2016ApJ...824...53C} provided definitive proofs for the presence of a highly ionised, relativistic wind in the source nuclear region. The strongly blueshifted absorptions of highly-ionised metals (i.e. Fe XXV, Si XIV) indicate that the outflow velocity components are within the range of $0.1-0.4~c$. The VLT/UVES spectrum of HS 0810+2554 also shows blueshifted absorptions of the C IV and N V doublets, indicating the existence of UV absorbing material moving with an outflowing speed of $v_{\rm C IV}=19,400$ km s$^{-1}$ \citepads{2014ApJ...783...57C,2016ApJ...824...53C}. Even though is  classified as radio-quiet object, VLA observations at 8.4 GHz \citepads{2015MNRAS.454..287J} indicate that HS 0810+2554 hosts a radio core, thus demonstrating that it is not radio-silent.

HS 0810+2554 was also recently observed with ALMA in the mm-band \citepads{2020MNRAS.496..598C}. The analysis of ALMA data has shown the tentative detection of high-velocity clumps of CO(J$=$3$\rightarrow$2) emission, suggesting the presence of a massive molecular outflow on kpc-scales. With our characterisation of the ionised outflow in HS 0810+2554, we now have, for the first time ever, a three-phase description of an AGN-driven wind at high redshift, from the nuclear to the galaxy scale: the highly-ionised (on nuclear scales), ionised, and neutral molecular phases (on galaxy scales) thanks to a broadband spectral coverage ranging from the X-ray to the optical and mm bands.
\subsubsection{SDSS J1353+1138} \label{sec:J}
Unlike HS 0810+2554, which has been widely observed in several spectral bands, SDSS J1353+1138 has been less intensively studied, as its discovery is more recent \citepads{2006ApJ...653L..97I}. This object was selected from the Sloan Digital Sky Survey (SDSS) as candidate double lensed QSO at $z\sim1.6$. \citetads{2006ApJ...653L..97I}  during the  University of Hawai'i 88-inch Telescope (UH88) follow-up observations of SDSS J1353+1138, obtaining \textit{V}, \textit{R}, \textit{I,} and \textit{H}-band images of the source. The two lensed images are well-distinguishable (see the \textit{right} panel in Fig. \ref{fig:opt_images}), with an angular separation of $\Delta \sim 1.40''$\citepads{2006ApJ...653L..97I}.

More recently, on 2016 January 13, SDSS J1353+1138 was observed with \textit{XMM-Newton} . The analysis of the X-ray spectrum \citep{Chartas.high-z.UFOs} revealed a significant absorption at $\sim6.8$ keV (consistent with Fe XXV), indicating the presence of a $\sim0.31c$ UFO.

\subsection{SINFONI observations and data reduction}
SINFONI observations of HS 0810+2554 and SDSS J1353+1138 were carried out on two different nights in February and March 2019, respectively, in the near-IR \textit{J}-band ($\lambda \sim 1.1-1.4$ $\mu$m) and with a spectral resolution $R=2000$. The observations were performed in seeing-limited mode\footnote{The SINFONI adaptive optics module (AO-mode) was not available at the time of the observations, since SINFONI had been moved from UT4 to UT3 for the last few months of its research activity.}, using the $0.250''\times0.125''$ pixel scale which provides a total field of view (FOV) of $8''\times8''$, which is essential for mapping the gas dynamics on galaxy scales. The airmasses are different for each target, spanning the ranges of $\sim1.7-1.9$ and $\sim1.2-1.3$ during the observations of HS 0810+2554 and SDSS J1353+1138, respectively. The data were obtained in eight and sixteen integrations of 300s each, for a total of 40 min for HS 0810+2554, and 80 min for SDSS J1353+1138. During each observing block, an ABBA pattern was followed: the target was put alternatively in two different positions of the FOV about $4.3''$ apart, to perform the sky subtraction through a nodding technique. A dedicated star observation to measure the point-spread-function (PSF) was not available in either case but the estimated angular resolution is $\sim$0.7$''$ ($\sim$0.8$''$) for HS 0810+2554 (SDSS J1353+1138), based on the measured extent of the spatially unresolved BLR emission (see Sect. \ref{sec:test_spatially_res}). Finally, a standard B-type star for telluric correction and flux calibration was observed shortly before or after the on-source exposures.

We used the ESO-SINFONI pipeline (v. 3.2.3) to reduce the SINFONI data. Before flux calibration and co-addition of single exposure frames, we corrected for atmospheric dispersion effects consisting in a significant change of the AGN continuum emission across the FOV of both sources. This is a consequence of the differential atmospheric dispersion at different wavelengths, which makes it so the measured spectra not ‘straight’ as expected. In practical terms, as the wavelength increases, an increasingly larger fraction of the emission gets deposited into adjacent pseudo-slits, producing a coherent spatial shift of the target as a function of $\lambda$. Additionally, possible flexures in the instrument may contribute to producing the observed shift.
In order to limit the impact of these optical distortions, we spatially aligned the emission centroid channel-by-channel in each single-exposure, sky-subtracted cube by adopting the following procedure. 

For each cube, we first determined, in every spectral channel, the average position of the emission centroid on the FOV through a 2D-Gaussian fit. Then we calculated the shift of the centroid mean position with respect to the centroid position in the first spectral channel, assumed as the reference channel. In both spatial directions on the FOV we found an increasing trend of the shift at increasing wavelengths, which we modelled with a two-degree polynomial to neglect the presence of some spikes that are due to noisier channels. The spatial shift, totally observed from the bluest to the reddest spectral channel, spanned the range of $\sim0.5-1$ pixel among the various single-exposure cubes of both targets. As the spatial shifts were fractional in units of pixels, we adopted the Drizzle algorithm \citep{drizzle_book, drizzle_book_bis} to perform the optimised alignment of every spectral channel in each raw single-exposure data cube.

Finally, we performed the flux calibration and the co-addition of the single-exposure cubes. The final sky-subtracted, flux-calibrated data consist of $100\times72\times2234$ data cubes, hence, each one including more than 7,000 spectra. Each spectrum is sampled by 2234 channels with a $\SI{1.25}{\angstrom}$ channel width and covers the spectral range $\sim1.1-1.4$ $\mu$m, corresponding to about $4400-5600$ $\SI{}{\angstrom}$ rest-frame wavelengths.

\subsection{Lens models for the two quasars} \label{sec:lens_model}
As both objects are gravitationally lensed QSOs, lens models are required in order to infer the intrinsic (i.e. unlensed) physical properties of the outflow, such as the intrinsic radius and unlensed flux, which are key ingredients for calculating the outflow energetics.

Both QSOs are lensed by a foreground elliptical galaxy and detailed lens models for these two objects can be found in the literature. In particular, for HS 0810+2554, there are several lens models reported in literature obtained from observations performed in different spectral bands (e.g. from VLA-radio data in \citeads{2015MNRAS.454..287J}, from ALMA-mm data in \citeads{2020MNRAS.496..598C}). In this work, we adopted, for HS 0810+2554, the most recent model by \citetads{2020MNRAS.492.5314N}, inferred from \textit{HST}-WFC3 IR observations: images and lens galaxy positions have been measured from direct F140W wide imaging (central wavelength $\sim$1392 nm), while slitless dispersed spectra have been provided by the grism G141 (useful range: $1075-1700$ nm). Assuming $z_{\rm l}\sim0.89$ for the lens galaxy \citepads{2011ApJ...738...96M}, \citetads{2020MNRAS.492.5314N}
modelled the deflector mass distributions with a singular isothermal ellipsoid (SIE), plus an external shear to account for tidal perturbations from nearby objects.

Detailed lens models for SDSS J1353+1138 are presented in \citetads{2006ApJ...653L..97I}  and \citetads{2016MNRAS.458....2R}, based on imaging observations in the \textit{i}-band with the Magellan Instant Camera (MagIC) on the Clay 6.5m Telescope and in the \textit{K}-band with the Subaru Telescope adaptive optics system, respectively. \citetads{2006ApJ...653L..97I} modelled the lens mass distribution using either a SIE model, or a singular isothermal sphere (SIS) model plus a shear component ($\gamma$), and estimated the lens redshift to be $z_{\rm l}\sim0.3$ based on the Faber-Jackson relation \citepads{1976ApJ...204..668F}. The resulting total magnification factors $\mu$ are 3.81 and 3.75 from the SIS+$\gamma$ and SIE model, respectively. Also assuming $z_{\rm l}\sim0.3$, \citetads{2016MNRAS.458....2R} found slightly lower values for total magnification: $\mu \sim3.47$ (SIS+$\gamma$),  $\mu \sim3.42$ (SIE) and $\mu \sim3.53$ (SIE+$\gamma$). We used all these magnification values from the literature to determine the unlensed flux carried by the ionised outflow in SDSS J1353+1138. Instead, for HS 0810+2554, we were able to provide an estimate of the ionised outflow magnification ($\mu_{\rm out}\sim2$) starting from our data. Such values will be  discussed further in Sect. \ref{sec:intrinsic} and Appendix \ref{app:2D_rec}.

\section{Data analysis} \label{sec:analysis}
\subsection{Fitting procedure}
For the spectral analysis of SINFONI data, we adopted the fitting code used in \citetads{2020A&A...644A..15M} to analyse MUSE data of two local QSOs. Here, we implemented the code to also handle the SINFONI data, introducing adjustments and new functionalities depending on the specific necessities of our data. In the following, we illustrate the basics of our spectral-fitting method, highlighting the required changes for the analysis of our SINFONI data (see \citeads{2020A&A...644A..15M} for the detailed description of the fitting code).

The entire fitting procedure aims at performing the kinematical analysis of the diffuse ionised gas, with a primary focus on the [O III]$\lambda$5007 emission line (hereafter [O III]) that is the optimal tracer of ionised outflows, as previously mentioned. Our strategy consists of the following three key-steps.

 \textit{Phase I.} We built a template model for the bright BLR emission using an integrated, high signal-to-noise ratio (S/N) spectrum.

 \textit{Phase II.} The BLR template built in phase I was used to map spaxel-by-spaxel the contribution of the BLR to the emission across the entire FOV. The resulting BLR model cube was then subtracted from the data cube.
 
 \textit{Phase III.} Finally, we performed, spaxel-by-spaxel, a finer modelling of the faint emission lines produced by the diffuse ionised gas, originating on galactic scales. 

Hereafter, we refer to the diffuse gas emission lines as ‘narrow’ in order to distinguish them from the typical ‘broad’ emission lines (FWHM $>$ 1000 km s$^{-1}$; e.g. \citeads{1981ApJ...249..462O}) originating from within the dense and highly turbulent BLRs.

A single noise value has been associated to each channel in our SINFONI data cubes, computed as the root mean square (rms) of the fluxes extracted spaxel-by-spaxel in a region with no significant emission from the target. The details of the spectral analysis of SINFONI data of both QSOs are provided below.

\captionsetup[subfigure]{labelformat=empty}
\begin{figure*}
\centering
  \includegraphics[width=17cm,keepaspectratio]{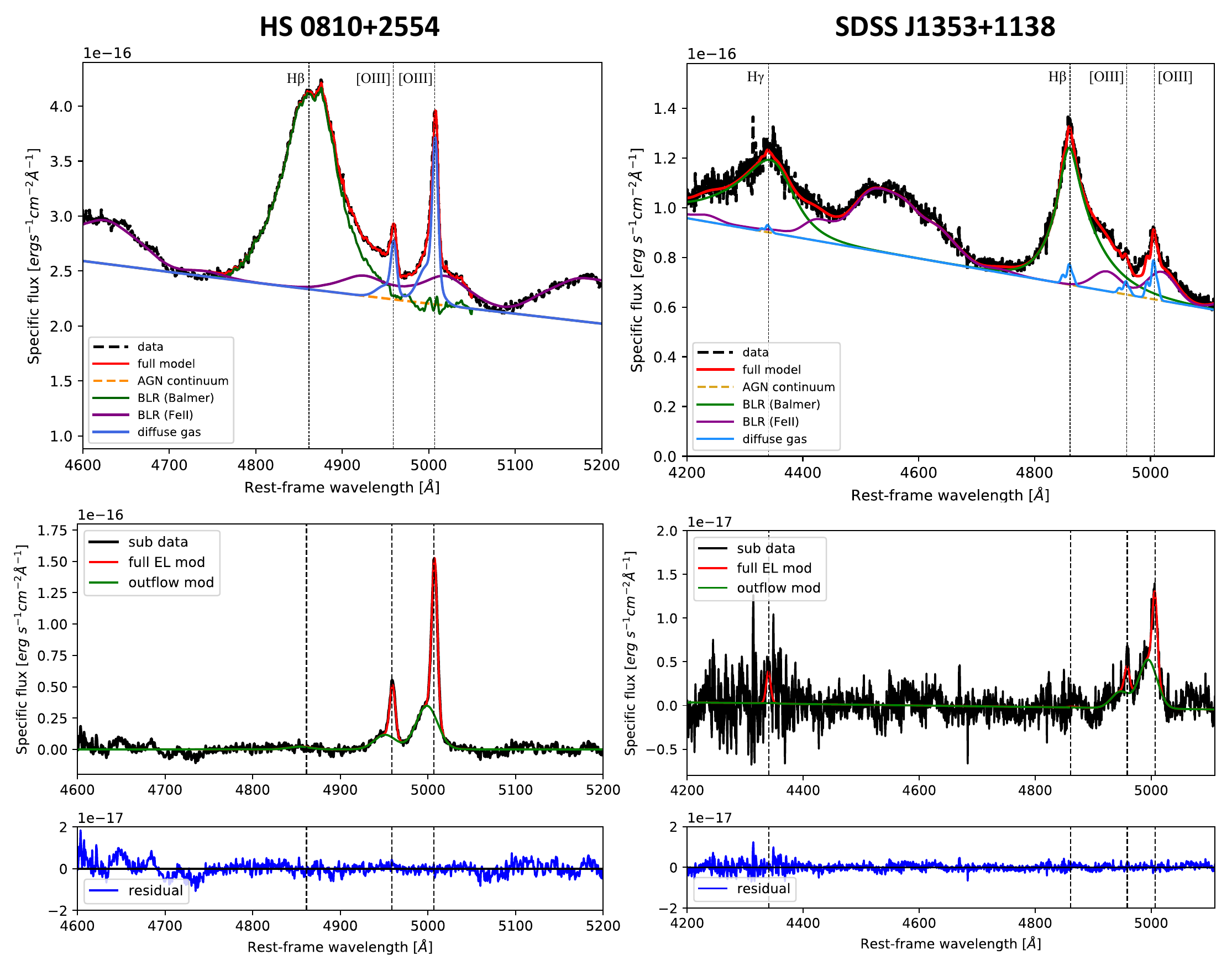}  
\caption{{\small Representative scheme of our fitting procedure. \textit{Top panels:} SINFONI \textit{J}-band spectra of HS 0810+2554 (\textit{left}) and SDSS J1353+1138 (\textit{right}), zoomed in the spectral region of [O III] and Balmer hydrogen emission lines. Both spectra were extracted using an aperture of $\sim$0.44$''$-radius, centred on the peak of the AGN continuum emission (located on image A in SDSS J1353+1138). Black dashed lines show the data, while red solid lines show the best-fit models obtained in phase II of our analysis. The latter is partitioned into the contributions of various components: AGN continuum (yellow), BLR emission in Balmer hydrogen emission lines (green) and Fe II (purple), and narrow line emission from the diffuse gas (light blue) represented as sum of multiple Gaussian components. \textit{Middle panels:} \textit{J}-band spectra extracted from the subtracted cubes, with the same aperture used in the  top panels. Subtracted data (black lines) are compared to the best-fit models (red lines) resulting from the finer emission lines (EL) modelling implemented in phase III. The green lines highlight the outflow component alone. \textit{Bottom panels:} Residuals obtained by subtracting the full EL model from the subtracted spectra.}}
\label{fig:int_spectra}
\end{figure*}

\subsubsection{I. Modelling the BLR emission} \label{sec:BLR_fit}
The fitting code starts with modelling the bright BLR emission in a spectrum extracted from the nuclear region, while also fitting the other spectral components. The spectral components to be fitted are: the AGN continuum, BLR emission lines, and narrow emission lines from the diffuse gas. In principle, we  should also have the stellar continuum emission, but in our data the AGN continuum is entirely dominant. The BLR model is built by the code as the sum of two independent components: the broad Balmer hydrogen emission lines (H$\beta$ in HS 08010+2554, H$\beta$ and H$\gamma$ in SDSS J1353+1138) and several Fe II broad emission lines, which are the two main BLR components in the rest-frame range observed by SINFONI ($\lambda \sim4200-5600$ $\SI{}{\angstrom}$). The diffuse emission instead consists of the [O III] emission doublet and the narrow components of the Balmer hydrogen lines.

For HS 0810+2554, we extracted a high-S/N spectrum from an aperture of 0.3$''$ radius, centred on the observed blended emission of A+B images (see Fig. \ref{fig:opt_images}), and fitted all the previously mentioned components simultaneously. We modelled the AGN continuum through a 1st-degree polynomial. The Fe II emission lines were modelled using the semi-analytic templates of \citetads{2010ApJS..189...15K}, while the BLR component of H$\beta$ was fitted by two broad Gaussian components. The narrow emission lines were fitted through two Gaussian components. Given the complexity of the BLR-H$\beta$ line profile in HS 0810+2554, we additionally associated spatially unresolved residuals from the fit to this component, following the approach  detailed in \citetads{2020A&A...644A..15M}.

In case of SDSS J1353+1138, where the two lensed images (A and B in Fig. \ref{fig:opt_images}) are well-distinguishable and spatially resolved, matters were complicated by the fact that we observed a significant change between nuclear spectra extracted from the two distinct images. As a consequence, this prevented us from considering a single BLR template. The procedure followed for SDSS J1353+1138 is described separately in Sect. \ref{sec:double_BLR}.

\subsubsection{II. Mapping the unresolved BLR emission across the FOV} \label{sec:tot_fit}
As the BLR emission goes unresolved in our data, we expect that its observed spatial variation follows the PSF of our observations. Therefore, we allowed the BLR template obtained in phase I to change only in amplitude across the FOV and we proceeded to fit the whole data cubes with the software pPXF (\citeads{2017MNRAS.466..798C}). For the modelling of the narrow emission lines, we used multiple Gaussian components and adopted a statistical approach (a Kolgomorov-Smirnov test) to select, spaxel-by-spaxel, the minimal, optimal number of Gaussian components to aptly reproduce  the emission line profiles (see \citeads{2020A&A...644A..15M} for details). For both HS 0810+2554 and SDSS J1353+1138, we considered a number of Gaussian components ranging from 1 to 3, finding the latter optimal for reproducing the most complex line profiles. Then we subtracted spaxel-by-spaxel the BLR and AGN continuum emissions and, for each galaxy, we created a cube containing only the residual emission lines due to the diffuse gas. In the following, we refer to this cube as the ‘subtracted cube’.

\subsubsection{III. Modelling the narrow emission lines} \label{sec:line_fit}
In phase III, we focused on the finer modelling of the narrow emission lines that remained after the subtraction of the BLR and AGN continuum emission components. The only significant residual emission in our data is the [O III] emission doublet, while the narrow components of Balmer hydrogen lines are very weak and marginally resolved. Therefore, we modelled the residual narrow emission lines through multiple Gaussian components, adopting some reasonable constraints: the two emission lines of the [O III] doublet were fitted by imposing the same central velocity and velocity dispersion, with the intensity ratio $I(5007)/I(4959)$ fixed at 3 according to the theoretical expectations of the atomic theory; whereas, for the weak narrow components of Balmer hydrogen lines, we assumed the same [O III]$\lambda$5007 line profile shape and left the flux as a free parameter. This is a reasonable assumption as we expect that the narrow Balmer hydrogen and [O III] emission lines come from regions with the same gas kinematics.

Similarly to the fit in phase II, we considered a number of Gaussian components ranging from 1 to 3, using, as before, a statistical approach to determine spaxel-by-spaxel the optimal number of components required. In both QSOs, most of the line profiles are well reproduced by two Gaussian components: a narrow, bright component close to the systemic velocity, plus a broad blueshifted component to reproduce the [O III] blue wing observed in most of the FOV, which we identify with approaching outflow emission. In HS 0810+2554, the 3-Gaussian fit was selected in some spaxels to reproduce properly the faint but still visible red wing in the [O III] line profile, tracing the fainter outflow component receding from us. On the contrary, in SDSS J1353+1138, two Gaussian components are sufficient to reproduce the most complex line profiles, as we do not detect the [O III] red wing anywhere.

In order to study the physical and dynamical properties of the outflow emission, which is the main focus of this work, we had to properly identify  the [O III] emission due to the high-speed outflowing gas by disentangling its contribution to the [O III] emission line due to the gas bulk motion within the host galaxy. To do so, we adopted the same selection criterium used by \citetads{2020A&A...644A..15M}. For each Gaussian component used to reproduce the [O III] line profile in a given spaxel, we focused on the fraction of the total line flux contained in the line wings with a velocity shift $|v-v_{\rm peak}|$ larger than a certain threshold width $w_{\rm th}$, where $v_{\rm peak}$ is the peak velocity of the line in each spaxel. If the fraction of total flux in the line wings was higher than a given threshold $\tau$, the Gaussian component has been classified as a possible ‘outflow’ component, to be confirmed by the following kinematical analysis (Sect. \ref{sec:kin_analysis}); otherwise, it has been classified as a ‘narrow’ component, due to systemic gas motions in the host galaxy. We verified the decomposition in several representative spaxels to select the optimal threshold values. In HS 0810+2554, we used $\tau=0.5$ and $w_{\rm th}=300$ km s$^{-1}$: the Gaussian component reproducing the narrowest, brightest emission near the systemic velocity has been typically classified as narrow; while any additional Gaussian component used to model either the blue or the red wing in [O III] profile has been identified as outflow component. In SDSS J1353+1138, in order to get the expected classification we slightly relaxed the width threshold ($w_{\rm th}=250$ km s$^{-1}$).

Figure \ref{fig:int_spectra} summarises our strategy. The top panels show \textit{J}-band spectra of HS 0810+2554 and SDSS J1353+1138, extracted from SINFONI data cubes with an aperture of $\sim$0.44$''$-radius, centred on the peak of the observed emission (located on image A in SDSS J1353+1138). The best-fit models shown were obtained in phase II. Since no distinctions between a possibly broad outflow and narrow systemic components were made at this stage of the procedure, the diffuse gas model is plotted as sum of multiple components.
We notice that in HS 0810+2554, there is a faint, broad emission line at $\sim4700-4750$ $\SI{}{\angstrom}$ (rest-frame) present in the Fe II templates, which is not present in our data. However, this does not affect the overall fitting procedure, as the observed Fe II emission is reproduced well by the templates at all the other wavelengths. The middle panels show the spectra extracted from the subtracted cubes using the same aperture as above, along with the results from the finer multi-Gaussian fit of the diffuse gas emission lines (phase III).
In both QSO-subtracted spectra, the [O III] line profile exhibits a prominent, asymmetric blue wing that is already visible in the full spectra shown above. This strongly suggests the presence of high-speed outflowing material moving towards the observer, which we discuss in greater detail in Sect. \ref{sec:kin_analysis}.

\subsection{Modelling a ‘double’ BLR in SDSS J1353+1138} \label{sec:double_BLR}

\begin{figure}
\begin{subfigure}{1.0\textwidth}
        \includegraphics[width=0.45\columnwidth, keepaspectratio]{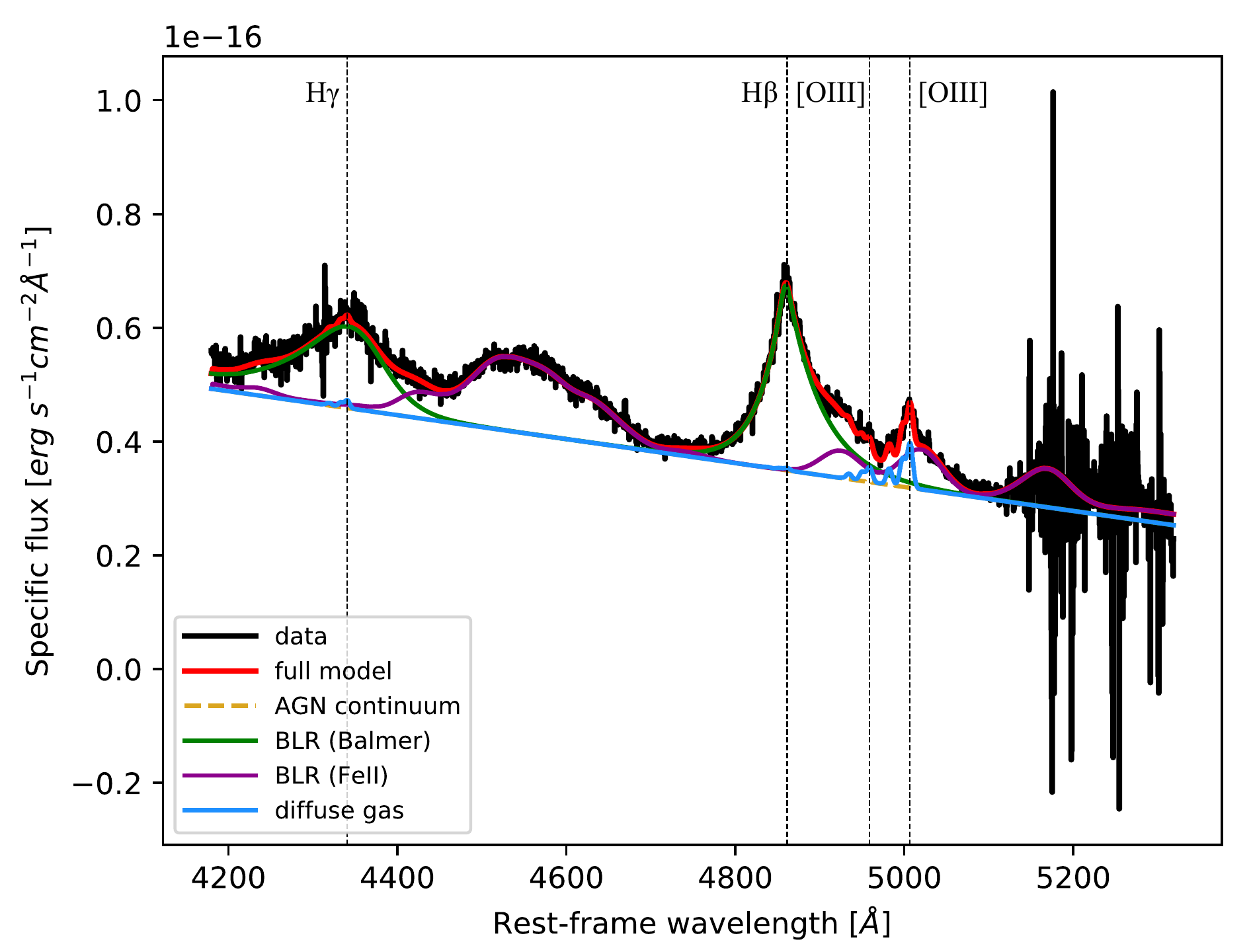}
        \caption{{\small}}
        \label{fig:J_blr_1}
\end{subfigure}
\newline
\begin{subfigure}{1.0\textwidth}
        \includegraphics[width=0.45\columnwidth, keepaspectratio]{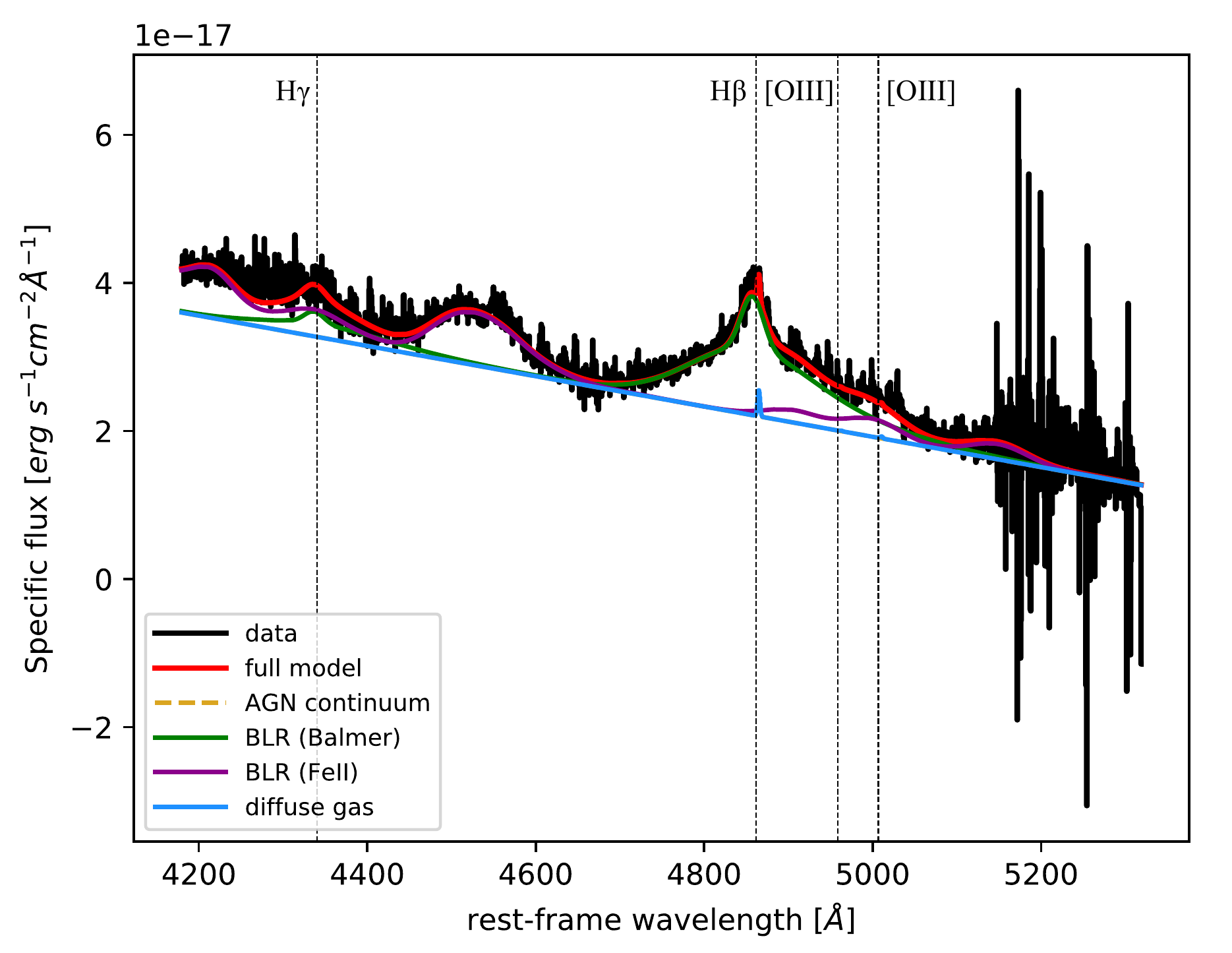}
        \caption{{\small}}
        \label{fig:J_blr_2}
\end{subfigure}
\caption{{\small Best-fit models of nuclear spectra of SDSS J1353+1138 extracted from an aperture of 0.3$''$ radius, centred on image A (upper panel) and on image B (lower panel). The various spectral components, the total model and data are represented with different colours (see the plot legend). In spectrum A, a broken power law distribution is perfectly suited to reproduce the BLR-H$\beta$ profile, while in spectrum B, an additional broad Gaussian component was required to adequately reproduce the broad peak in the BLR-H$\beta$ line profile, which is entirely dominant over the barely detected H$\beta$ narrow component (solid light-blue line). The two spectra clearly differ from each other mostly for the lack of [O III] detection and the presence of a prominent blue wing in the H$\beta$ line profile in spectrum B.}}
\label{fig:J_blr}
\end{figure}

As noted in Sect. \ref{sec:BLR_fit}, for SDSS J1353+1138, we found a significant change in the spectral shape within the wavelength range including the H$\beta$ and [O III] lines, while comparing the nuclear spectrum of the brighter image A (spectrum A) with that of the fainter image B (spectrum B), shown in the upper and lower panels of Fig. \ref{fig:J_blr}, respectively. In particular, while in the former it is possible to easily identify  the [O III] emission lines, we did not detect any counterpart in the latter. Moreover, the H$\beta$ line profile in spectrum B is broader, with an evident brighter blue wing. Both effects are likely due to an overall increase in the Fe II emission in image B, as the H$\beta$ line width is not expected to intrinsically vary between different lensed images.
The anti-correlation between Fe II and [O III] emissions in AGN spectra reflects a well-known effect known as Eigenvector-1 \citepads{1992ApJS...80..109B} and represents one of the most frequent differences among AGN properties. Even though it has been the subject of many studies, a clear physical understanding of its origin is still lacking. \citetads{1992ApJS...80..109B} suggest that high column densities in the BLR enhance Fe II, while reducing the ionising radiation able to reach the NLR. In a spectral analysis of AGN principal components in SDSS, \citetads{2009ApJ...706..995L}  argue instead that the covering factor of the NLR is the likely cause of the range in [O III] strength, while \citetads{2009ApJ...707L..82F} suggest that the higher column densities required for the infall in more luminous AGNs would additionally account for the observed correlation of Fe II strength with $L/L_{\rm Edd}$.

In spite of its still unclear origin, there are two main possible explanations for the observed significant variation in the Fe II emission between the two lensed images. The first one is based on the typical short time scales (i.e. days, weeks; e.g. \citeads{2000ApJ...533..631K}) on which the BLR is seen to vary. Because of the different path followed by the light from the background QSO, the two lensed images are produced with a time delay of about 16 days \citepads{2006ApJ...653L..97I}. This temporal shift is comparable with the typical BLR variation timescale, therefore, it could be sufficient to have a significant change in the BLR emission explaining the effect we observed. Given the short  time scale probed here, this could carry interesting implications for the accretion variations in the AGN and in the consequent response of the BLR gas. Alternatively, the observed variation could be the consequence of microlensing effects (e.g. see \citeads{2020MNRAS.492.5314N}) produced by either single stars or low-mass dark matter halos intervening along the line of sight. Microlensing effects typically affect only the emission originated on small scales, while the emission from the NLR is insensitive. Of the two possibilities, the latter seems to be less likely, as we do not observe any significant counterpart variation in the strength of the BLR-H$\beta$ component, in addition to that observed in the Fe II strength. However, a remarkable simultaneous variation in both H$\beta$ and Fe II strength is what we would expect in the case when the two emissions are strictly co-spatial, whereas we know that the BLR is stratified and that microlensing magnifies the emission from the most compact regions more strongly. Therefore, we do not exclude the microlensing hypothesis. A detailed analysis of the different BLR spectra from the two images is beyond the scope of this work and will be presented in a forthcoming paper. Therefore, in this work, we focus on how we accounted for this effect during the spectral analysis.

Consequently, in SDSS J1353+1138, we extracted two distinct nuclear spectra, namely, spectrum A and B that are shown in Fig. \ref{fig:J_blr}, using a 0.3$''$ radius aperture centred on the emission peak of each lensed image. We proceeded to fit them separately. In both spectra we used a first-degree polynomial to model the AGN continuum that is still dominant over the stellar continuum. Unlike the BLR modelling of HS 0810+2554, the multiple Gaussian fit was not sufficient to reproduce the broader and complex profile of the broad Balmer emission lines, especially the H$\beta$ line in spectrum B. In fact, even though both H$\beta$ prominent wings are likely due to the Fe II emission, as discussed above, the Fe II templates employed by the code were not able to reproduce such observed emission. Therefore, we modelled both wings as part of the broad H$\beta$ line profile without  any focus on their physical interpretation, as we were simply interested in identifying the overall BLR spectrum in order to remove it. For the modelling of the broad Balmer emission lines observed in SDSS J1353+1138 (i.e. H$\beta$ and H$\gamma),$  we used a broken power law distribution convolved with a Gaussian profile \citepads{2006A&A...447..157N}:
\begin{equation} \label{eq:broke_powlaw}
F(\lambda) = 
\begin{cases}
        F_0 \times \left( \frac{\lambda}{\lambda_0} \right )^{+\alpha}, &\text{for }\ \lambda < \lambda_0 \\ 
        F_0 \times \left( \frac{\lambda}{\lambda_0} \right )^{-\beta}, &\text{for }\ \lambda > \lambda_0
\end{cases}
,\end{equation}
where the free parameters of the fit, for each line, are the central wavelength $\lambda_0$, the two power-law indices $\alpha$ and $\beta$, the normalisation $F_{\rm 0}$, and the width $\sigma$ of the Gaussian kernel. In spectrum A (upper panel), the H$\beta$ and H$\gamma$ were modelled separately through the line profile described in Eq. (\ref{eq:broke_powlaw}). Spectrum B (lower panel)  required  instead an additional broad Gaussian component to suitably reproduce  both the more extended red wing and the broad peak ($\sigma \sim 800$ km s$^{-1}$) in the H$\beta$ profile. Moreover, we had to constrain the H$\gamma$ line profile through that of H$\beta$. Different Fe II templates have been selected in the BLR best-models for the two nuclear spectra. To model the narrow emission line profiles, we used three Gaussian components in spectrum A, while a single Gaussian component in spectrum B, since we  did not actually observe any narrow component.

Then we proceeded to fit the whole data cube following the procedure described in Sect. \ref{sec:tot_fit}, with the main difference that in each spatial pixel, pPXF considered a linear combination of the two BLR models, weighing their relative contribution and providing their most suitable combination as the BLR model in that specific spaxel. In general, in those spaxels close to one of the lensed images, we basically got  the BLR model directly obtained in the modelling of the respective nuclear spectrum; while a combination of the two BLR-models in those spaxels ended up roughly between the two images, as we indeed expected.

\subsection{Testing the spatially resolved emission of the ionised outflows} \label{sec:test_spatially_res}

Before analysing the outflow kinematics across the FOV, we tested whether the emission of the detected ionised outflows was spatially resolved. This is crucial for the calculation of the outflow energetics. As our observed data have missed a dedicated PSF star, we compared the spatial extent of the [O III] outflow emission with that of the BLR emission (both obtained from the previous spectral modelling). In fact, given that the latter is unresolved in our data, it is suitable for reproducing the instrumental response. We fitted a 2D-Gaussian profile to the BLR flux map, obtained by integrating in wavelength our BLR model, in order to estimate the angular resolution of our seeing-limited observations. The resulting best-fit Gaussian profiles are not circularly symmetric, especially in the case of HS 0810+2554, whose profile is elongated in the NW-SE direction. Such an elongation is mostly due to lens stretching effects and to the blending of A and B images, rather than to a possible intrinsic asymmetry of the PSF. Therefore, for both sources we assumed the angular size of the minor axis of the best-fit Gaussian profile as representative for the true PSF extent, as this is heading in the direction where the lens stretching effects are expected to be minimised. For HS 0810+2554 and SDSS J1353+1138, we estimated for the PSF a FWHM ($\theta_{\rm res}$) of 0.7$''$ and 0.8$''$, respectively. These NIR values are slightly smaller than the optical seeing measurements obtained with the differential image motion monitor (DIMM) during the observations, namely, $0.9''$ and $1.0''$, respectively \citepads{cite-key}.

\begin{figure*}
\centering
\includegraphics[width=17cm,keepaspectratio]{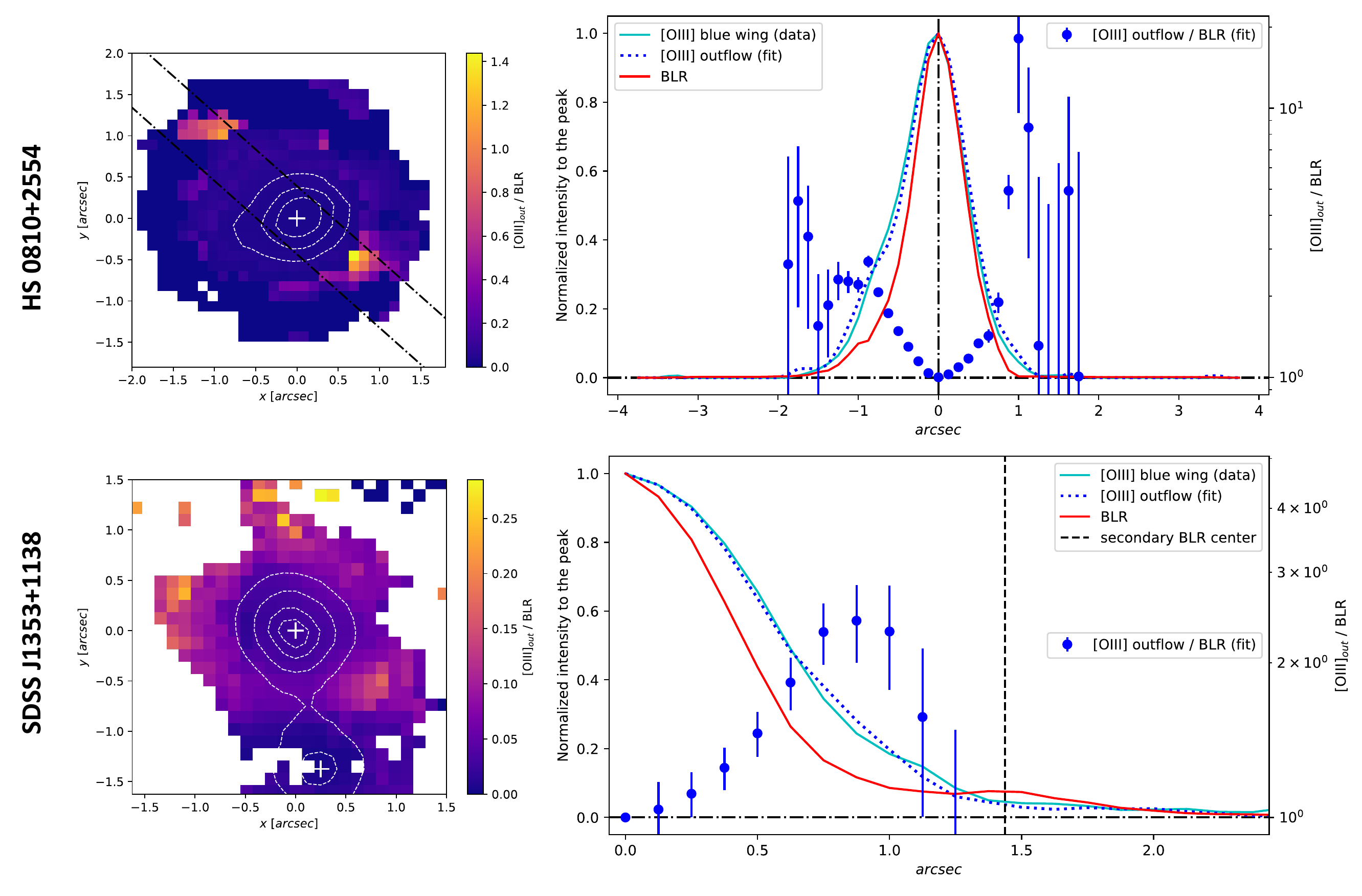}
\caption{{\small Spatially resolved ionised outflows in HS 0810+2554 (top) and in SDSS J1353+1138 (bottom). \textit{Left panels:} Maps of the ratio between the [O III] outflow flux and the BLR flux (both from best-fit models). Coloured pixels refer to a S/N $\gtrsim$ 2 (S/N $\gtrsim$ 3) on the full [O III] emission line (i.e. narrow + outflow components) for HS 0810+2554 (SDSS J1353+1138). The positions of the continuum emission peak of the lensed images are marked with white ‘+’, while the dotted white lines indicate the contour levels of the BLR emission at 75\%, 50\%, and 25\% of its peak; for SDSS J1353+1138, it is represented also the level at 90\%. \textit{Right panels:} Normalised intensity profiles along the pseudo-slit (black dotted-dashed lines in the ratio map) and in circular annuli of increasing radius for HS 0810+2554 (top) and SDSS J1353+1138 (bottom), respectively: BLR model (red lines), [O III] outflow model (dotted blue lines) and [O III] blue wing from data (cyan lines). Blue points represent ratio values of the [O III] outflow flux over the BLR flux and they are referred to the right-hand logarithmic scale. The dashed black line in the plot of SDSS J1353+1138 corresponds to the radial distance of the centre of image B.}}
\label{fig:test_resolved}
\end{figure*}

To test whether the detected ionised outflows are spatially resolved, we adopted the following procedure. First, we created the flux maps for the [O III] outflow and BLR components. Then we calculated spaxel-by-spaxel the ratio of the [O III] outflow ([O III]$_{\rm out}$) flux over the BLR flux, and produced the maps reporting the flux-ratio values across the FOV.

The ratio maps obtained for both QSOs are shown in the left panels of Fig. \ref{fig:test_resolved}. The increasing trend in [O III]$_{\rm out}$-to-BLR ratios with the distance from the emission peak indicates that the [O III] outflows are spatially resolved in both QSOs. Moreover, the ratio map of HS 0810+2554 highlights the existence of a preferred NE-SW direction along which the highest ratio values are found. Such a direction is perpendicular to the blending direction of the images A+B. Unlike HS 0810+2554, the two lensed images of SDSS J1353+1138 are  spatially well-resolved and not affected by significant lens stretching effects. As a consequence, the [O III]$_{\rm out}$-to-BLR ratio maps shows an isotropic pattern of increasing ratio values moving outwards from the centre of image A (we recall that we detected the [O III] emission only from this image, as previously discussed in Sect. \ref{sec:double_BLR}).

What we discuss above, based on the use of 2D-ratio maps, can be better appreciated using spatial profiles. We determined the spatial profiles of the [O III] outflow and BLR emissions, as well as of their ratio values and studied their variability with increasing distance from the peak of the overall emission. In HS 0810+2554, since the [O III] emission is preferentially located along the NE-SW direction, we defined a pseudo-slit in such a direction ($\theta\sim130^{\circ}$) with a width of five spaxels, along which we calculated the emission spatial profiles. On the contrary, given the isotropic pattern of the whole emission in SDSS J1353+1138, we determined the spatial profiles in circular annuli of increasing radial distance from the centre of image A (lower panel). All the spatial profiles thus obtained are shown in the right panels of Fig. \ref{fig:test_resolved}. Each one has been normalised to its own $0''$-value, that is, to the value at the peak position of the overall emission; in the case of the [O III] outflow and BLR profiles, the $0''$-value corresponds also to their own peak value. This is helpful in further confirming our previous conclusion that the [O III] outflows are spatially resolved in both galaxies since the [O III] outflow profiles are broader than the respective BLR profiles. A unique exception occurs in SDSS J1353+1138, in a correspondence of about 1.5$''$ from the centre where we can observe a clear bump in the BLR emission profile: this is due to the flux contribution from image B. Therefore, this does not affect our previous conclusion. As a further test, in addition from that obtained from our fitting procedure, we determined the [O III] outflow profile by collapsing the spectral channels in the subtracted-data cube, including the blue wing of the [O III] line profile ($4976-5000$ $\SI{}{\angstrom}$ and $4970-4996$ $\SI{}{\angstrom}$ for HS 0810+2554 and SDSS J1353+1138, respectively). The two [O III] outflow profiles, from the fit (blue dotted line) and from the spectrally-integrated (cyan solid line) subtracted-data, agree very well. 

In order to estimate the angular extent of ionised outflows on the image plane\footnote{To be still corrected for lens effects.}, we focused on the ratio values of the [O III] outflow flux over the BLR flux. These are plotted in logarithmic scale (on the right-hand side of the plots), after having been rescaled to 1 in the central pixel. In this way, we can easily identify values higher than 1 as regions producing a significant [O III] outflow emission and, hence, we can determine the spatial extent of resolved ionised outflows. The associated errorbars were computed by propagating the uncertainties on the [O III] and BLR fluxes in the spatial pixels involved. These were computed by propagating the error (mostly due to the noise) associated to the spectral channels, over which we integrated to get the total flux contained in that spatial pixel. We took the maximum distance, including solely ratio values not consistent with 1, as both radius within which to spatially integrate the flux of the [O III] outflow component, and outflow observed radius ($R_{\rm obs}$) to be still corrected for the SINFONI-PSF and lensing stretching effects. To correct for the PSF-smearing, we applied the correction $R_{\rm PSF}= \sqrt{R^2_{\rm obs}-(\theta_{\rm res}/2)^2}$, where $R_{\rm PSF}$ is the PSF-corrected radius, $R_{\rm obs}$ is the radius observed in the image plane and $\theta_{\rm res}$ is our seeing estimate (0.7$''$ and 0.8$''$ for HS 0810+2554 and SDSS J1353+1138, respectively). In HS 0810+2554, we spatially integrated the [O III] outflow flux up to $R_{\rm obs}=1.25''$, finding a total observed flux of $(3.73\pm0.05)\times10^{-15}$ erg s$^{-1}$ cm $^{-2}$ and $R_{\rm PSF}\sim1.2''$. In SDSS J1353+1138, we took $R_{\rm obs}=1.13''$ and assessed a total observed flux of $(8.6\pm0.6)\times10^{-16}$ erg s$^{-1}$ cm$^{-2}$ and $R_{\rm PSF}\sim1.06''$\footnote{The estimates for $R_{\rm PSF}$ are here provided with no uncertainty. We evaluate the error on the outflow intrinsic radius in Sect. \ref{sec:intrinsic}, after correcting for the lensing effects.} for the [O III] outflow. In Sect. \ref{sec:intrinsic}, we account for the lens effects and estimate the intrinsic extent (by correcting for stretching effects) and unlensed flux (by correcting for magnification effects) of the ionised outflows in HS 0810+2554 and in SDSS J1353+1138.

\section{Results} \label{sec:results}
\subsection{Distribution and kinematics of the ionised gas} \label{sec:kin_analysis}

\begin{figure*}
\centering
\includegraphics[width=17cm,keepaspectratio]{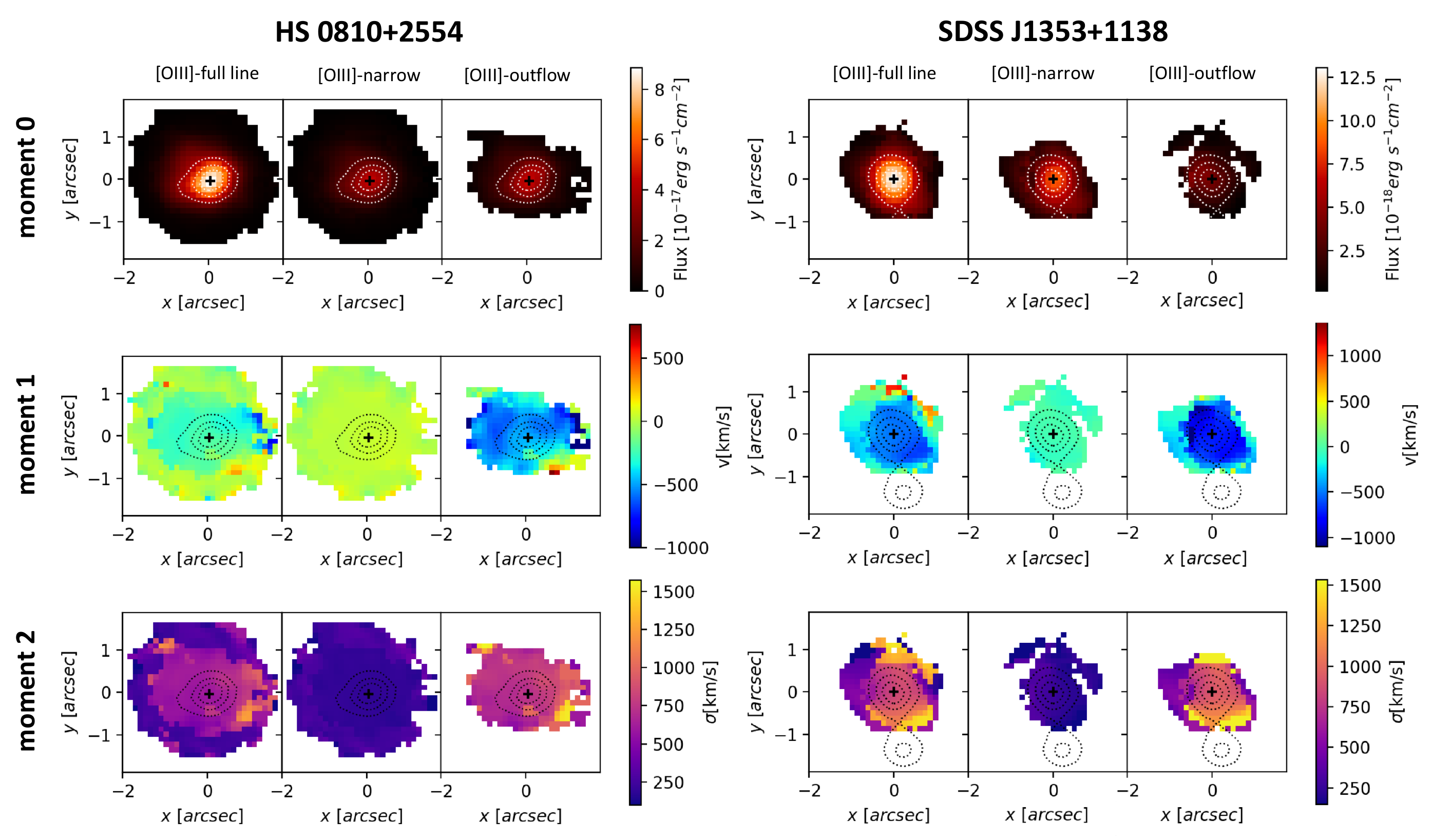}
\caption{{\small Moment-0 (intensity field), moment-1 (velocity field), and moment-2 (dispersion field) maps of the [O III] line emission in HS 0810+2554 (\textit{left}) and in SDSS J1353+1138 (\textit{right}). The maps for the total, narrow and outflow components are shown separately, reporting only spatial pixels with a S/N equal or higher than 2 for HS 0810+2554, and than 3 for SDSS J1353+1138. The black ‘+’ indicates the emission centroid in each QSO, while the dotted lines represent the contour levels of the BLR emission at 75\%, 50\%, and 25\% of its peak.}}
\label{fig:kin_maps}
\end{figure*}

The main purpose of this work is to map the kinematics of the [O III] emission, with a primary focus on the outflow component. Figure \ref{fig:kin_maps} shows a global overview of the distribution and the kinematics of the ionised gas resulting from the modelling of the [O III] emission line. The moment-0 (intensity field), moment-1 (velocity field), and moment-2 (dispersion field) maps for the narrow and the outflow components are shown separately in order to better trace their distinct spatial and velocity distributions. All maps have been produced reporting only spatial pixels with a S/N equal or higher than 2 for HS 0810+2554 and higher than 3 for SDSS J1353+1138.

The candidate [O III]-outflow component is extended up to large distance from the galaxy centre of both QSOs, and it stands out for its strongly-blueshifted velocities and high velocity dispersion values ($|v|\gtrsim500$ km s$^{-1}$ and $\sigma \gtrsim 600$ km s$^{-1}$, respectively; see moment-1 moment-2 maps in Fig. \ref{fig:kin_maps} relative to the outflow component). Such velocity dispersions are well above the values measured in typical star-forming systems at these redshifts ($\sigma \sim 100$ km s$^{-1}$,  e.g. \citeads{2009ApJ...697..115C,2009ApJ...697.2057L}) and, along with the overall blue-shifted motion, they provide clear evidence for large-scale outflows in these galaxies. Moreover, while in HS 0810+2554 the outflow and the narrow components have almost the same intensity across the FOV, we note that in SDSS J1353+1138, the [O III] outflow emission is brighter than the [O III] narrow emission produced by the bulk of the gas of the host galaxy.

For the narrow component, which is expected to trace mostly systemic galactic motions, we obtained low-velocity and low velocity-dispersion values ($|v|\lesssim50$ km s$^{-1}$, $\sigma \lesssim200$ km s$^{-1}$ for HS 0810+2554, and $|v|\lesssim100$ km s$^{-1}$, $\sigma \lesssim300$ km s$^{-1}$ for SDSS J1353+1138; see moment-1 moment-2 maps in Fig. \ref{fig:kin_maps} relative to the narrow component) in the central region where also the outflow emission is detected, further supporting our decomposition in the spectral analysis of both QSOs. In the more external region of HS 08101+2554, we observe slightly higher values in the velocity dispersion ($\sigma \lesssim300$ km s$^{-1}$), as the line profile is modelled with a unique Gaussian component, given that the [O III] emission line is fainter and the S/N is lower. This could indicate that the [O III] outflow component is still present, but cannot be isolated from the [O III] narrow component because of its faintness and the worse S/N.

\begin{figure}
\resizebox{\hsize}{!}{\includegraphics{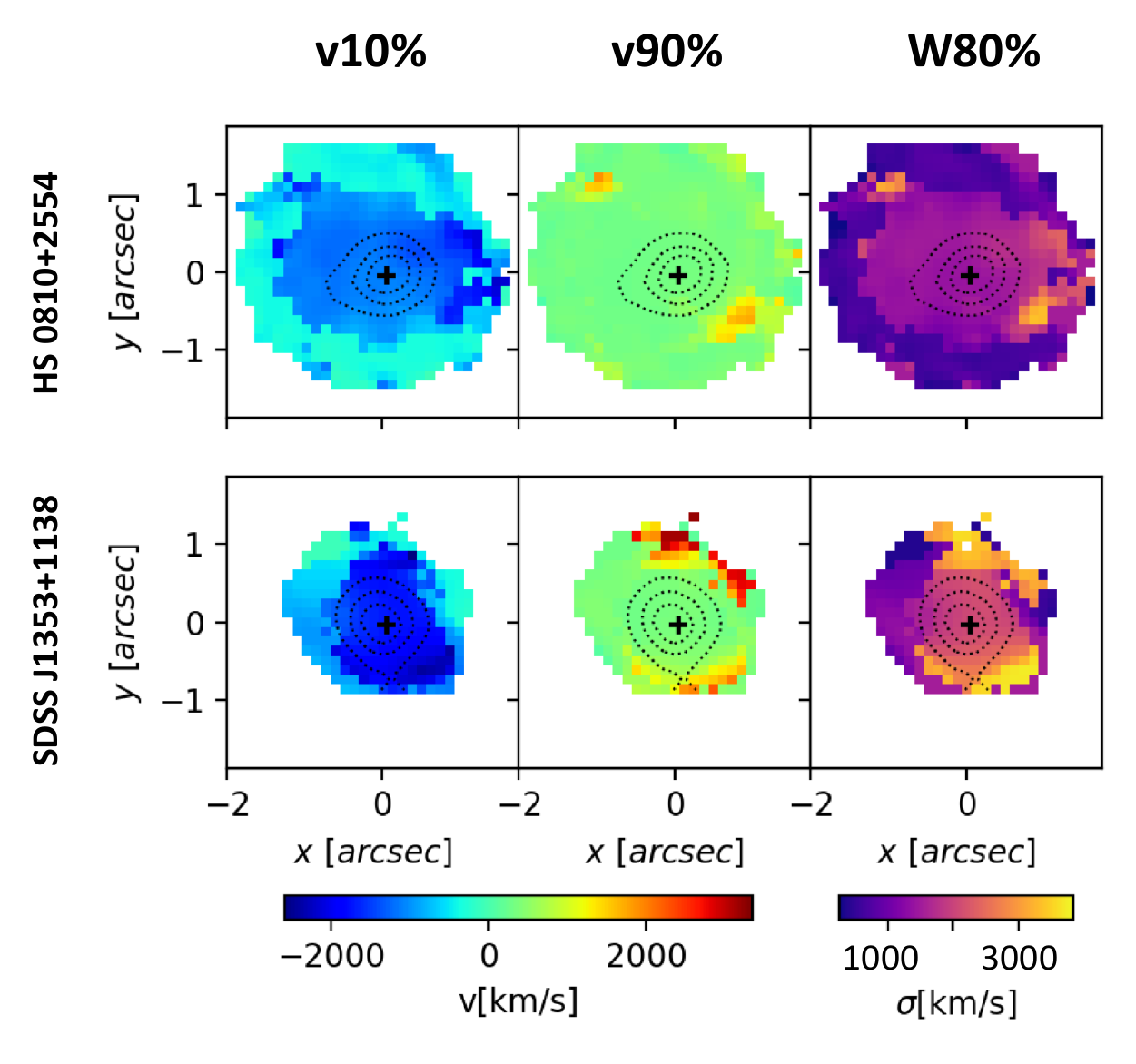}}
\caption{{\small $v_{\rm10}$, $v_{\rm90}$, and $W_{\rm80}$ maps of the [O III] emission line in HS 0810+2554 (upper panels) and in SDSS J1353+1138 (lower panels). We applied the same cut in S/N as in the moment maps of Fig. \ref{fig:kin_maps}, that is, with a S/N equal or higher than 2 and for HS 0810+2554 and higher than 3 for SDSS J1353+1138.}}
\label{fig:perc_map}
\end{figure}

Similarly to \citetads{2020A&A...644A..15M} and given the complexity of the [O III] line profile across the FOV, we preferred adopting the following definitions of velocity and width for the outflow characterisation (e.g. see also \citeads{2014MNRAS.441.3306H, 2014MNRAS.442..784Z, 2015ApJ...799...82C, 2015A&A...580A.102C, 2016A&A...588A..58B}), rather than the moment-1 and moment-2 values. The latter are indeed more affected by geometrical projection and dust absorption effects. In each spatial pixel, we determined the 10th and 90th velocity percentiles ($v_{\rm 10}$ and $v_{\rm 90}$) of the overall emission line profile (i.e. narrow + outflow components if present), as representative velocities of the approaching and receding outflow components, respectively. The null velocity value corresponds to the systemic velocity peak of the narrow component in the central spectrum. From $v_{\rm 10}$ and $v_{\rm 90}$, we computed the line width $W_{\rm 80}$ defined as $v_{\rm90}-v_{\rm10}$. The $W_{\rm80}$ width is approximately equal to the full width at half maximum (FWHM) for a Gaussian profile.  Maps of $v_{\rm10}$, $v_{\rm90}$ and $W_{\rm80}$ are shown in Fig. \ref{fig:perc_map}.

The maps of $v_{\rm10}$ show highly blueshfited velocities in most of the field of HS 0810+2554 and SDSS J1353+1138. In the former, a slightly steeper velocity gradient is present in the west direction from the centre, where we observe velocities as high as about $-2170$ km s$^{-1}$; in the latter, the outflow region is preferentially elongated in the NE-SW direction with highest velocity values (up to $-2410$ km s$^{-1}$) at the SW end of the strongly blueshifted region. In HS 0810+2554, we clearly detect also the redshifted component of the outflow in the two reddest regions in the $v_{\rm90}$ map, where the outflow is seen receding from us at velocities up to about $1730$ km s$^{-1}$ along the line of sight. Looking at the $W_{\rm80}$ maps, we observe extremely large values ($1100~\text{km s$^{-1}$}\lesssim v \lesssim 3500$ km s$^{-1}$) in the outflow regions, which is consistent with other $z\sim2$ QSO outflows found in the literature \citepads{2015A&A...580A.102C}. Furthermore, by comparing the [O III]-outflow moment-1 map with the $v_{\rm10}$ map for each QSO, we note that the shape of the [O III]-outflow moment-1 map reflects the bluest region in the $v_{\rm10}$ map, suggesting that any additional Gaussian component added to model the wings in the [O III] profile has been correctly classified as outflow component (compare also [O III]-outflow moment-2 maps and the respective $W_{\rm80}$ maps).

We rule out the possibility of alternative scenarios, such as galactic inflows or a galaxy merger event. In fact, in the few reported cases of their detection, galactic inflows have been observed mostly in absorption and with quite small bulk velocities ($\sim200$ km s$^{-1}$) and velocity dispersions (e.g. \citeads{2013Sci...341...50B}). Moreover, for the inflowing gas theoretical modelling predicts a small covering factor (e.g. \citeads{2010ApJ...717..289S}), making its direct observation rare especially at high redshift (e.g. \citeads{2010Natur.467..811C}). Finally, we exclude also the galaxy-merger scenario since the deep optical images of both HS 0810+2554 and SDSS J1353+1138 (see Fig. \ref{fig:opt_images}) do not show any continuum emission counterpart in correspondence of the outflow region, which could support such a scenario.

Finally, we stress that the obtained maps are relative to the lens plane and, thus, they do not account for gravitational lensing effects. While these are expected not to significantly affect the observed gas kinematics (hence the outflow velocity), they strongly alter the observed gas spatial distribution and the observed surface brightness: fluxes are magnified and spatial dimensions are stretched. Therefore, the obtained maps could not be used to infer directly the outflow intrinsic radius and its total flux, which are key ingredients, along with the outflow velocity, in the computation of the outflow energetics. We first need  to quantify the lensing effects and then we can derive the unlensed physical properties of the outflow. This aspect is discussed in Sect. \ref{sec:intrinsic} (see also Appendix \ref{app:2D_rec}).

\subsection{Inferring unlensed size and flux of the outflow} \label{sec:intrinsic}

As we discuss in Sect. \ref{sec:kin_analysis}, we performed both the spectral analysis and the kinematical study in the lens plane. For this reason, we had to correct for the lens effects to determine the actual extent and flux of the outflow.

There are several adaptive-mesh fitting codes which, given a mass distribution for the lens and a surface brightness profile for the background source, fit the resulting forward lensed image to the observed data and use a statistics test (e.g. the minimum $\chi^2$ method) to establish the best-fit models for both the lens and the source. These algorithms usually require the knowledge of the instrumental PSF to allow a correct comparison with the observed data. The output of these fitting-codes is a 2D or 3D reconstruction (depending on the code used) of the unlensed source. In order to achieve an accurate reconstruction, it is required that the lensed images are all detected and spatially resolved\footnote{In addition or alternatively to single lensed images, fitting codes handle also lensed arcs.}, as their position depends on the first derivative of the gravitational potential of the lens, while their flux on the second derivative (e.g. \citeads{2015MNRAS.454..287J, 2020MNRAS.492.5314N}). In other words, the knowledge of the position of the multiple lensed images and of their flux provides strong constraints on the lens and background source models.

Unfortunately, we could not use
such fitting codes to fully reconstruct the unlensed outflow in the source
plane for  HS 0810+2554 nor SDSS J1353+1138 since our data did not satisfy the necessary requirements. In fact, in the case of HS 0810+2554 the spatial resolution of the SINFONI data was too low to resolve the various lensed images and, thus, we were not able to achieve an accurate full reconstruction of the background source. However, we were able to recover a partial reconstruction of the background outflow using the lens-fitting code from \citetads{2018MNRAS.481.5606R} and adopting an approximated procedure (see Appendix \ref{app:2D_rec}). In this way, we estimated the outflow magnification factor and intrinsic radius to be, respectively, $\mu_{\rm out}=2.0\pm0.2$ and $R_{\rm out}=(8.7\pm1.7)$ kpc, with $z=1.508\pm0.002$ as the redshift measured from the nuclear spectrum extracted for the BLR-modelling (described in Sect. \ref{sec:BLR_fit}) and, here, adopted to convert the outflow angular size into kpc units. Our $z$ measurement is consistent with the centroids of the ALMA CO(J$=$3$\rightarrow$2) and CO(J$=$2$\rightarrow$1) emission lines of HS 0810+2554 reported in \citetads{2020MNRAS.496..598C}. By correcting for $\mu_{\rm out}$ the observed [O III] outflow flux (determined in Sect. \ref{sec:test_spatially_res}), we found the unlensed outflow flux to be $F_{\rm out}=(1.9\pm0.2)\times10^{-15}(2.0/\mu_{\rm out})$ erg s$^{-1}$cm$^{-2}$. Our $\mu_{\rm out}\sim2$ estimate is close to what \citetads{2020MNRAS.496..598C} found for a high-velocity CO-clump at similar distance in ALMA data of HS 0810+2554. On the contrary, it remarkably differs (up to two orders of magnitude) from the values from the literature, presented in Sect. \ref{sec:HS}. This follows from the fact that the latter are estimates of the magnification of the emission from the more compact (UV disk and X-ray corona) region, while our estimate is relative to a large-scale ($\sim$8 kpc) emission. Moreover, the reconstructed unlensed outflow does not intercept the lens caustics (see Appendix \ref{app:2D_rec}) and, hence, it misses the magnification contribution from those regions, where the lens magnification is drastically larger.

In SDSS J1353+1138, the complete lack of [O III] detection in image B prevented us from attempting any background source reconstruction starting from our data, as no constraints could be put on this image. Therefore, we had to make simplified assumptions, referring to the lens models for SDSS J1353+1138 by \citetads{2006ApJ...653L..97I} and \citetads{2016MNRAS.458....2R} (discussed in Sect. \ref{sec:lens_model}), based on AGN plus host galaxy emission in the \textit{i} and \textit{K}-band, respectively. Our assumptions rely on the fact that in SDSS J1353+1138 gravitational lensing effects are supposed to be smaller than in HS 0810+1154 (e.g. \citeads{1996astro.ph..6001N}). As a consequence, as compared to HS 0810+1154, for this object we expect: (1) a milder and almost isotropic stretching of physical dimensions, as we indeed observed; (2) smaller and less variable-in-space values of differential magnification. On the basis of the first argument, we neglected the stretching lens effect and approximated the unlensed outflow angular size to $R_{\rm PSF}=1.06''\pm0.13''$ (determined in Sect. \ref{sec:test_spatially_res}). Considering instead the lens magnification, we expect total magnification factors of a few units that is weakly dependent on the geometrical details of the flux distribution for background emissions with comparable spatial extent. Consequently, we used the average between the \textit{i}-band (i.e. $\mu=3.81$ and $\mu=3.75$; \citeads{2006ApJ...653L..97I}) and \textit{K}-band total magnification factors (i.e. $\mu=3.47$, $\mu=3.42$ and $\mu=3.53$; \citeads{2016MNRAS.458....2R}) as a proxy for the total outflow magnification $\mu_{\rm out}$, under the assumption of comparable unlensed physical sizes.
Given the unknown real unlensed flux distribution of the \textit{J}-band outflow, we conservatively assumed an uncertainty of 10\% on our adopted $\mu_{\rm out}$ value, thus obtaining $\mu_{\rm out}=3.6\pm0.4$. Correcting, finally, for the lens magnification and converting to kpc-units, we found the unlensed flux for the outflow to be $F_{\rm out}=(2.4\pm0.3)\times10^{-16}(3.6/\mu_{\rm out})$ erg s$^{-1}$cm$^{-2}$ and its intrinsic radius to be $R_{\rm out}=(9.2\pm1.1)$ kpc, using $z=1.632\pm0.002$ as measured from the nuclear spectra extracted during the BLR-modelling (described in Sect. \ref{sec:double_BLR}).

In Table \ref{tab:kin_properties}, we summarise the main outflow properties for HS 0810+2554 and SDSS J1353+1138, obtained up to this point. The first columns show the maximum velocity values observed in the $v_{\rm10}$, $v_{\rm90}$ and $W_{\rm80}$ maps (described in Sect. \ref{sec:kin_analysis}), respectively, referred to as $v^{\rm max}_{\rm10}$, $v^{\rm max}_{\rm90}$ and $W^{\rm max}_{\rm80}$. Then we report our lens-corrected estimates of $R_{\rm out}$ and $F_{\rm out}$ (inferred as discussed above), the latter corrected for the outflow magnification factors, $\mu_{\rm out}$, shown in the last column. Most of these physical quantities are also employed in the computation of the outflow energetics in Sect. \ref{sec:energetics}.

\begin{table*}
\centering
\begin{tabular}{c|cccccc}
\hline
\hline
  QSO & $v^{\rm max}_{\rm10}$ & $v^{\rm max}_{\rm90}$ & $W^{\rm max}_{\rm80}$ & $R_{\rm out}$ & $F_{\rm out}$ & $\mu_{\rm out}$\\
    &  km s$^{-1}$ & km s$^{-1}$ & km s$^{-1}$ & kpc & erg s$^{-1}$cm$^{-2}$ & \\
 \hline
 HS 0810+2554 & $-2170\pm70$ & $1730\pm60$ & $3360\pm110$ & $8.7\pm1.7$ & $(1.9\pm0.2)\times10^{-15}$ & $2.0\pm0.2$\\ 
 SDSS J1353+1138 & $-2410\pm80$ & $2270\pm130$ & $3850\pm90$ & $9.2\pm1.1$ & $(2.4\pm0.3)\times10^{-16}$ & $3.6\pm0.4$\\ 
  \hline
\end{tabular}%
\caption{{\small Directly measured properties of the [O III] outflows in HS 0810+2554 and SDSS J1353+1138, obtained from our analysis, and adopted outflow magnification factors. Starting from left, columns are defined as follows: maximum velocity values observed in the $v_{\rm10}$, $v_{\rm90}$ and $W_{\rm80}$ maps ($v^{\rm max}_{\rm10}$, $v^{\rm max}_{\rm90}$ and $W^{\rm max}_{\rm80}$, respectively), intrinsic outflow radius ($R_{\rm out}$), unlensed [O III] outflow flux ($F_{\rm out}$) corrected for the outflow magnification factor ($\mu_{\rm out}$), reported in the last column.}}
\label{tab:kin_properties}
\end{table*}%

\subsection{Outflow energetics} \label{sec:energetics}
We derived the physical properties of the large-scale ionised outflows in HS 0810+2554 and SDSS J1353+1138 from the observed [O III]$\lambda$5007 emission, following the prescriptions described in \citetads{2012A&A...537L...8C} as done also in \citetads{2020A&A...644A..15M}. The [O III] line luminosity is given by:
\begin{equation}\label{eq:oiii_line_lum}
L_{\rm [O III]}=\int_V \epsilon_{\rm [O III]}f~dV,
\end{equation}
where $V$ is the volume occupied by the ionised outflow, $f$ is the filling factor of the [O III] emitting clouds in the outflow, and $\epsilon_{\rm [O III]}$ is the [O III] line emissivity which, at the temperatures typical of the NLR ($\sim10^4$ K), is weakly dependent on the temperature ($\propto T^{0.1}$) and can be written as:
\begin{equation}\label{eq:line_emissivity}
\epsilon_{\rm [O III]}=1.11\times10^{-9}E_{\rm [O III]}n_{\rm O^{2+}}n_{\rm e}~\text{erg s}^{-1}\text{cm}^{-3},
\end{equation}
with $E_{\rm [O III]}$ as the energy of the [O III] photons, $n_{\rm O^{2+}}$ and $n_{\rm e}$, the volume densities of the $\rm O^{2+}$ ions, and of the electrons, respectively. Then assuming that most of the oxygen in the ionised outflow is in the form of $\rm O^{2+}$, it follows that:
\begin{equation}\label{eq:line_emissivity_approx}
\epsilon_{\rm [O III]}\approx5\times10^{-13}E_{\rm [O III]}n^2_{\rm e}10^{\rm [O/H]}~\text{erg s}^{-1}\text{cm}^{-3},
\end{equation}
where $\rm [O/H]$ gives the oxygen abundance in solar units. The mass of the outflowing ionised gas can be derived from the following expression:
\begin{equation}\label{eq:mass_out_int}
M_{\rm out}=\int_V 1.27m_{\rm H}n_{\rm e}f~dV,
\end{equation}
where $m_{\rm H}$ is the mass of the hydrogen atom and the factor of 1.27 follows from including the mass contribution of helium.
By combining Eqs. (\ref{eq:oiii_line_lum}) and (\ref{eq:mass_out_int}), we get:
\begin{equation}\label{eq:mass_out}
\resizebox{0.9\columnwidth}{!}{%
        $M_{\rm out}=5.33\times10^7~\bigg(  \frac{L_{\rm [O III]}}{10^{44}~\text{erg s}^{-1}} \bigg)~\bigg(  \frac{\langle n_{\rm e} \rangle}{10^3~\text{cm}^{-3}} \bigg)^{-1}~\frac{C}{10^{\rm [O/H]}}~\text{\(M_\odot\)},$
        }
\end{equation}
\\
where $\langle n_{\rm e} \rangle$ is the electron density averaged over the ionised outflow volume (i.e. $\langle n_{\rm e} \rangle=\int_V n_{\rm e}f~dV/\int_Vf~dV$) and $C=\langle n_{\rm e} \rangle^2/\langle n^2_{\rm e} \rangle$ is the so-called ‘condensation factor’. Under the simplifying hypothesis that all ionising gas clouds have the same density, we get $C=1$ and eliminate the outflow mass dependance on the filling factor of the emitting clouds.

In order to compute the energetics of the ionised outflow, we have to make further simplifying assumptions for the outflow geometry and structure: we assume that the outflow has a (bi-)conical geometry with an opening angle $\Omega$ and a radial extent, $R_{\rm out}$, and that it consists of a collection of ionised gas clouds, uniformly distributed within the cone and outflowing with a speed $v_{\rm out}$. The mass outflow rate is given by:
\begin{equation}\label{eq:mdot_def}
\dot{M}_{\rm out}=\langle \rho \rangle~v_{\rm out}~\Omega R^2,
\end{equation}
where $\langle \rho \rangle$ is the average mass density computed over the total volume $V$ occupied by the conical outflow\footnote{We note that unless $f=1$, in general $\langle \rho \rangle \neq 1.27m_{\rm H}\langle n_{\rm e} \rangle$, with $\langle n_{\rm e} \rangle$ averaged over the volume occupied by the emitting clouds and not over the whole conical volume.}. By substituting $\langle \rho \rangle$ in Eq. (\ref{eq:mdot_def}) with its definition in terms of $\dot{M}_{\rm out}$ (using Eq. \ref{eq:mass_out}) and $V$, we obtain:
\\
\begin{equation}\label{eq:mdot}
        \resizebox{0.9\columnwidth}{!}{%
        $\dot{M}_{\rm out}=164\bigg(  \frac{L_{\rm [O III]}}{10^{44}\text{erg s}^{-1}} \bigg)\bigg(  \frac{\langle n_{\rm e} \rangle}{10^3\text{cm}^{-3}} \bigg)^{-1}\bigg(  \frac{v_{\rm out}}{10^3\text{km s}^{-1}} \bigg)\bigg(  \frac{R_{\rm out}}{\text{kpc}} \bigg)^{-1}\frac{\text{\(\text{M}_\odot\)yr}^{-1}}{10^{\rm [O/H]}}$
        }
\end{equation}
\\
where we have assumed $C=1$. The mass outflow rate thus inferred does not depend on either the opening angle $\Omega$ of the outflow cone or the filling factor $f$ of the emitting clouds.

Finally, we calculate the kinetic energy ($E_{\rm kin}$), kinetic luminosity ($L_{\rm kin}$) and momentum rate ($\dot{p}_{\rm out}$) of the outflow by means of the following expressions:
\\
\begin{equation}\label{eq:kin_energy}
E_{\rm kin}=9.94\times10^{42}~\bigg(  \frac{M_{\rm out}}{\text{\(\text{M}_\odot\)}} \bigg)~\bigg(  \frac{v_{\rm out}}{\text{km s}^{-1}} \bigg)^2~\text{erg}
,\end{equation}
\begin{equation}\label{eq:kin_lum}
L_{\rm kin}=3.16\times10^{35}~\bigg(  \frac{\dot{M}_{\rm out}}{\text{\(\text{M}_\odot\)yr}^{-1}} \bigg)~\bigg(  \frac{v_{\rm out}}{\text{km s}^{-1}} \bigg)^2~\text{erg s}^{-1} 
,\end{equation}
\begin{equation}\label{eq:pdot}
\dot{p}_{\rm out}=6.32\times10^{30}~\bigg(  \frac{\dot{M}_{\rm out}}{\text{\(\text{M}_\odot\)yr}^{-1}} \bigg)~\bigg(  \frac{v_{\rm out}}{\text{km s}^{-1}} \bigg)~\text{dyne}.
\end{equation}
\\
Equations (\ref{eq:mass_out})-(\ref{eq:pdot}) require the knowledge of different physical properties of the outflow, some of which we were able to derive, while others had to be assumed. These are: the oxygen abundance, which we fixed to the solar abundance, and the electron density, which we assumed to be $n_{\rm e}\sim1000$ cm$^{-3}$, in agreement with the values measured in similar studies at high redshift (see e.g. \citeads{2017A&A...606A..96P, 2019ApJ...875...21F}). The second assumption, in particular, affects the derived outflow energetics \citepads{2020MNRAS.498.4150D, 2020A&A...642A.147K} but it is necessary, nonetheless, since we cannot measure $n_{\rm e}$ directly from our data.

We now focus on the physical quantities we were able to calculate. In Sect. \ref{sec:intrinsic}, we provided the values of the intrinsic radius $R_{\rm out}$ and flux $F_{\rm out}$ of the ionised outflows, traced by the [O III] line emission. From $F_{\rm out}$ we calculated the intrinsic [O III] line luminosity used in the mass outflow expression (Eq. \ref{eq:mass_out}).
We found ($\mu_{\rm out}$-corrected) [O III] luminosity values of $L_{\rm [O III]}=(2.8\pm0.3)\times10^{43}$ erg s$^{-1}$ and $L_{\rm [O III]}=(4.4\pm0.6)\times10^{42}$ erg s$^{-1}$ for HS 0810+2554 and SDSS J1353+1138, respectively.

In order to establish the velocity of the ionised outflows, we focused on the kinematical maps of $v_{\rm 10}$ and $v_{\rm 90}$, shown in Sect. \ref{sec:kin_analysis}. The spectral analysis and the study of the gas kinematics in HS 0810+2554 and in SDSS J1353+1138 have revealed the presence of an extended central region hosting outflowing gas directed towards us along the line of sight (Sect. \ref{sec:kin_analysis}), leading to very high values of $v_{\rm 10}$. Moreover, in HS 0810+2554, we detected also a red wing in the [O III] line profile in two smaller regions apart from the peak of the overall emission, corresponding to the high-velocity receding component of the ionised outflow. In this case, following \citetads{2020A&A...644A..15M}, we defined the outflow velocity as:
\begin{equation}\label{eq:v_out_HS}
v_{\rm out}=max~(|v^{\rm max}_{\rm 10}-v_{\rm sys}|,~ |v^{\rm max}_{\rm 90}-v_{\rm sys}|)
,\end{equation}
where $v^{\rm max}_{\rm 10}$ and $v^{\rm max}_{\rm 90}$ are the maximum value, respectively, observed in the $v_{\rm 10}$ and $v_{\rm 90}$ maps (described in Sect. \ref{sec:kin_analysis}), and $v_{\rm sys}$ is the bulk (or systemic) velocity of the galaxy, inferred from the nuclear spectrum used for the BLR-fitting (in Sect. \ref{sec:BLR_fit}) and set to the value of 0 km s$^{-1}$, as previously described in Sect. \ref{sec:kin_analysis}. This definition is required by the unknown geometry and orientation of the outflow with respect to the line of sight: since we ignore the true angle of the outflow with respect to the line of sight, and given that the bulk of the outflow unlikely points towards the observer, we assume that the best representation of the outflow speed is provided by the velocity ‘tail’ of the line profile, that is, $v_{\rm 10}$ and $v_{\rm 90}$ in Eq. (\ref{eq:v_out_HS}). These values are thought to be more suited to represent $v_{\rm out}$ than the mean (or median) velocity of the line, which strongly depends on projection effects and dust absorption (e.g. \citeads{2015ApJ...799...82C}).

The same argument holds also for the determination of the outflow velocity in SDSS J1353+1138. But in this case, possibly because of the different orientation with respect to the observer and higher dust absorption, we did not observe any asymmetric red wing in the [O III] profile produced by the receding part of the outflow, as stressed in Sect. \ref{sec:kin_analysis}. Therefore, for SDSS J1353+1138 we focused only on the blue tail of the [O III] line associated to the outflow, and hence assumed $v_{\rm out}=v^{\rm max}_{\rm 10}$ as outflow velocity for SDSS J1353+1138.

To calculate the quantities in Eqs. (\ref{eq:mass_out})-(\ref{eq:pdot}) and their uncertainties, we used the error propagation considering the inferred errors on $R_{\rm out}$, $L_{\rm [O III]}$ and $v_{\rm out}$, and a typical uncertainty of 50\% on $n_{\rm e}$ (e.g. \citeads{2017A&A...606A..96P, 2019ApJ...875...21F}). The physical properties of the ionised outflows detected in HS 0810+2554 and in SDSS J1353+1138 are reported in Table \ref{tab:energetics}, including also our estimates of the kinetic efficiency and of the momentum-boost. The former is defined as the ratio between the outflow kinetic luminosity, $L_{\rm kin}$, (defined in Eq. \ref{eq:kin_lum}) and the AGN bolometric luminosity, $L_{\rm Bol}$, (corrected for the lens magnification), while the latter is defined as the ratio between the momentum rate of the outflow ($\dot{p}_{\rm out}$) and the momentum initially provided by the AGN-radiation pressure (i.e. $L_{\rm Bol}/c$), which is approximately identified also with the momentum rate of the X-ray UFO. The values of $L_{\rm Bol}$ adopted in this work are for $(2.5\pm0.9)\times10^{45}$ erg s$^{-1}$ and $(39\pm2)\times10^{45}$ erg s$^{-1}$ HS 0810+2554 and SDSS J1353+1138, respectively, and they will be discussed in Sect. \ref{sec:connecting}.

We found mass outflow rates of $\sim$2 \(\text{M}_\odot\)yr$^{-1}$ and $\sim$12 \(\text{M}_\odot\)yr$^{-1}$, and kinetic efficiencies of $\sim9\times10^{-5}$ and $\sim700\times10^{-5}$ for SDSS J1353+1138 and HS 0810+2554, respectively. The values obtained for HS 0810+2554 are in good agreement with the predictions at $L_{\rm Bol}\sim2\times10^{45}$ erg s$^{-1}$ of the $\dot{M}_{\rm out}-L_{\rm Bol}$ and $L_{\rm kin}-L_{\rm Bol}$ scaling relations \citepads{2015A&A...580A.102C, 2017A&A...601A.143F} for the ionised outflow component. On the contrary, the inferred $\dot{M}_{\rm out}$ value for SDSS J1353+1138 is small (by a factor of $\sim100$) compared to the predictions for the ionised outflow component at $L_{\rm Bol}\sim4\times10^{46}$ erg s$^{-1}$. Our findings are however relative only to the ionised phase of large-scale outflows, while a significant amount of outflowing gas may be in neutral molecular and/or atomic phase. Given the typical bolometric luminosities of our galaxies ($L_{\rm Bol}\sim10^{45}-10^{46}$ erg s$^{-1}$), outflow mass rates predicted for the molecular component may indeed exceed our measurements for the ionised gas by a factor $\sim100$. \citetads{2020MNRAS.496..598C} have recently claimed the tentative detection of a massive ($M_{\rm out,~mol}\sim4\times10^{9}$ \(\text{M}_\odot\)) CO-molecular outflow in HS 0810+2554, with total mass rate and velocity of $\sim$400 \(\text{M}_\odot\)yr$^{-1}$ and 1040 km s$^{-1}$, respectively. Compared to molecular outflows, which have been observed in $z>1$ QSOs in spite of the uncertainty due to detection limits (e.g. \citeads{2017A&A...605A.105C, 2017A&A...608A..30F, 2018A&A...612A..29B}), neutral atomic outflows mainly reside in ultra-luminous infrared galaxies (ULIRGs), showing both intense SF and AGN activity (e.g. \citeads{2005ApJS..160..115R, 2016A&A...590A.125C, 2017A&A...606A..96P, 2019MNRAS.483.4586F, 2020arXiv200613232F}). This may suggest that neutral atomic outflows are mostly powered by SF rather than by AGN activity (e.g. \citeads{2017A&A...606A..36C, 2018ApJ...853..185B}) or that they can occur in obscured AGN hosting large quantities of cold gas to be channeled into galaxy-scale outflows \citepads{2017A&A...606A..96P}. However, the contribution of any additional gas phase may significantly increase the overall outflow mass and energetics (e.g. \citeads{2014A&A...562A..21C, 2016A&A...591A..28C, 2017A&A...601A.143F}). The implications of all these results will be discussed in details in Sect. \ref{sec:discussion}.

\begin{table*}
\centering
\begin{tabular}{c|cccccc}
\hline
\hline
  QSO & $v_{\rm out}$ & $M_{\rm out}$ & $\dot{M}_{\rm out}$ & $E_{\rm kin}$ & $L_{\rm kin}/L_{\rm Bol}$ & $\dot{p}_{\rm out}$\\
    & \text{km s$^{-1}$} & $10^6$ \(\text{M}_\odot\) & \(\text{M}_\odot\)yr$^{-1}$ & $10^{56}$ erg & $10^{-5}$ & $L_{\rm Bol}/c$\\
 \hline
 HS 0810+2554 & $2170\pm70$ & $15\pm8$ & $12\pm6$ & $7\pm4$ & $700\pm500$ & $1.9\pm1.2$\\ 
 SDSS J1353+1138 & $2410\pm80$ & $2.4\pm1.2$ & $1.9\pm1.0$ & $1.4\pm0.7$ & $9\pm7$ & $0.022\pm0.018$\\ 
  \hline
\end{tabular}%
\caption{{\small Derived properties of the ionised outflow in HS 0810+2554 and SDSS J1353+1138, as derived from the analysis of the [O III] line emission. From the left, columns report for the ionised outflows the measured values of: velocity, mass, mass rate, kinetic energy, kinetic efficiency and momentum-boost. The values of $L_{\rm Bol}$ are shown in Table \ref{tab:energetics_UFOs}.}}
\label{tab:energetics}
\end{table*}%

\section{Discussion} \label{sec:discussion}

\subsection{Connection with the nuclear X-ray UFOs} \label{sec:connecting}

The main purpose of this work is to shed light on the acceleration mechanism of ionised outflows on large scales. In this regard, we compared the energetics of the galaxy-scale ionised outflows to the UFOs present at nuclear scales, in order to test whether they are causally connected (i.e. whether they are subsequent phases of the same AGN-accretion burst). In the case of a causal connection, we could go on to investigate the nature of the acceleration mechanism distinguishing between momentum-driven and energy-driven winds.

We show the energetics measurements of the large scale ionised outflows in HS 0810+2554 and SDSS J1353+1138 in Sect. \ref{sec:energetics}. In this section, we present the X-ray measurements of the hosted UFOs from the literature that are used to determine the sub-pc scale wind energetics. For both QSOs, we refer to \citet{Chartas.high-z.UFOs}, who present the first X-ray spectral analysis of SDSS J1353+1138, and new UFOs measurements for HS 0810+2554 obtained from a new \textit{Chandra} observation acquired in 2016 using the updated version of the photoionisation code \textit{XSTAR} \citepads{2001ApJS..134..139B}.

As done in \citetads{2020A&A...644A..15M}, we followed \citetads{2018MNRAS.478.2274N} in order to re-compute the UFOs energetics in a consistent way based on the same assumption we made in the calculation of the large-scale outflow energetics. We assumed that the UFO is launched from the escape radius $r_{\rm esc}\equiv2GM_{\rm BH}/v^2_{\rm UFO}$ of the BH and we derived the mass outflow rate for the nuclear wind as:
\begin{equation}\label{eq:mdot_UFO}
\dot{M}_{\rm UFO}\simeq0.3~ \bigg( \frac{\Omega}{4\pi} \bigg) \bigg(  \frac{N_{\rm H}}{10^{24}\text{cm}^{-2}} \bigg)\bigg(  \frac{M_{\rm BH}}{10^8\text{\(\text{M}_\odot\) yr}^{-1}} \bigg)\bigg(  \frac{v_{\rm UFO}}{c} \bigg)^{-1} \text{\(\text{M}_\odot\)yr}^{-1}
,\end{equation}
where $\Omega$ is the solid angle subtended by the UFO, $M_{\rm BH}$ the black hole mass, $N_{\rm H}$ the hydrogen gas column density, and $v_{\rm UFO}$ the wind speed. We took the values of $v_{\rm UFO}$ and $N_{\rm H}$ inferred by \citet{Chartas.high-z.UFOs}. For HS 0810+2554, these values (reported in Table \ref{tab:energetics_UFOs}) slightly differ from those previously published (i.e. $v_{\rm UFO,1}=0.12^{+0.02}_{-0.01}c$, $N_{\rm H,1}=3.4^{+1.9}_{-2.0}\times10^{23}~\text{cm}^{-3}$ and $v_{\rm UFO,2}=0.41^{+0.07}_{-0.04}c$, $N_{\rm H,2}=2.9^{+2.0}_{-1.6}\times10^{23}~\text{cm}^{-3}$ for the two distinct UFO components; \citeads{2016ApJ...824...53C}), but are still in agreement. The adopted $M_{\rm BH}$ values are virial estimates based on H$\beta$ for HS 0810+2554 \citepads{2011ApJ...742...93A} and on C IV for SDSS J1353+1138 \citep{Chartas.high-z.UFOs}. The C IV-based measurement of $M_{\rm BH}$ for SDSS J1353+1138 has been corrected following the prescription published in \citetads{2017MNRAS.465.2120C} for C IV-based virial black hole mass estimates, which are known to be systematically biased compared to masses derived from the Balmer hydrogen lines. Finally, we assumed for both UFOs a covering factor of $f=\frac{\Omega}{4\pi}=0.4^{+0.2}_{-0.2}$, on the basis that about 40\% of local AGNs have been observed to host UFOs \citepads{2010A&A...521A..57T, 2013MNRAS.430...60G}. With all these ingredients, we calculated the mass rate $\dot{M}_{\rm UFO}$ (with Eq. \ref{eq:mdot_UFO}), the momentum rate $\dot{p}_{\rm UFO}$ (i.e. $\dot{p}_{\rm UFO}=\dot{M}_{\rm UFO}v_{\rm UFO}$) and the momentum-boost of the nuclear winds defined as $\dot{p}_{\rm UFO}/(L_{\rm Bol}/c)$. As uncertainty on all quantities, the minimum-maximum range of possible values was considered, considering the values of $v_{\rm UFO}$, $N_{\rm H}$ and $M_{\rm BH}$ as inferred in \citet{Chartas.high-z.UFOs}. For $L_{\rm Bol}$ we took the average between the $\mu$-corrected values obtained in \citet{Chartas.high-z.UFOs} with two different methods: the 2-10 keV bolometric correction \citepads{2012MNRAS.425..623L} and the estimate from the continuum luminosity at 1450 $\SI{}{\angstrom}$ \citepads{2011ApJ...742...93A, 2012MNRAS.422..478R}.

Table \ref{tab:energetics_UFOs} summarises the physical properties of the UFOs in HS 0810+2554 and SDSS J1353+1138, derived following the prescriptions in \citetads{2018MNRAS.478.2274N}, along with the physical quantities taken from \citet{Chartas.high-z.UFOs}. For HS 0810+2554 the values of $\dot{M}_{\rm UFO}$ and $\dot{p}_{\rm UFO}/(L_{\rm Bol}/c)$ refer to the whole hosted UFO, as sum of the two detected UFO components at different speeds (for which we separately report $N_{\rm H}$ and $v_{\rm UFO}$ in Table \ref{tab:energetics_UFOs}), originally discovered in \citetads{2016ApJ...824...53C} and confirmed in \citet{Chartas.high-z.UFOs}.
\begin{table*}
\centering
\setlength{\tabcolsep}{6pt} 
\renewcommand{\arraystretch}{1.3} 
\begin{tabular}{c|ccccc|cc}
\hline
\hline
   QSO  & log($M_{\rm BH}$)$~^{\rm a}$ & $L_{\rm Bol}~^{\rm b}$ & $N_{\rm H}$ & $v_{\rm UFO}$ & $f$ & $\dot{M}_{\rm UFO}$ & $\dot{p}_{\rm UFO}$\\
    & \(\text{M}_\odot\) & $10^{45}$ erg s$^{-1}$ & $10^{23}$ cm$^{-3}$ & $c$ &  & \(\text{M}_\odot\)yr$^{-1}$ & $L_{\rm Bol}/c$\\
 \hline
 \multirow{2}{*}{HS 0810+2554} & \multirow{2}{*}{$8.62^{+0.22}_{-0.22}$} & \multirow{2}{*}{$2.5\pm0.9$} & $2.1^{+1.0}_{-1.1}$ & $0.11^{+0.05}_{-0.03}$ & \multirow{2}{*}{$0.4\pm0.2$} & \multirow{2}{*}{$1.2^{+4.2}_{-1.1}$} & \multirow{2}{*}{$3.9^{+17.6}_{-3.5}$}\\ 
  &  &  & $1.4^{+0.3}_{-0.5}$ & $0.43^{+0.04}_{-0.05}$ &  &  & \\ 
 \hline
 SDSS J1353+1138 & $9.41^{+0.30}_{-0.30}$ & $39\pm2$ & $3.9^{+2.4}_{-2.3}$ & $0.34^{+0.02}_{-0.09}$ & $0.4\pm0.2$ & $4^{+9}_{-3}$ & $1.7^{+9.7}_{-1.6}$\\ 
   \hline
\end{tabular}%
\caption{{\small Physical properties of the UFOs in HS 0810+2554 and SDSS J1353+1138, as derived from the X-ray measurements reported in \citet{Chartas.high-z.UFOs} by using the prescription of  \citetads{2018MNRAS.478.2274N}. (a) The $M_{\rm BH}$ values are virial estimates based on H$\beta$ in HS 0810+2554 \citepads{2011ApJ...742...93A}, and on C IV in SDSS J1353+1138 \citep{Chartas.high-z.UFOs}, respectively. The C IV-based measurement of $M_{\rm BH}$ in SDSS J1353+1138 has been corrected following the prescription for C IV-based virial black hole mass estimates published in \citetads{2017MNRAS.465.2120C}. (b) The values of $L_{\rm Bol}$ are corrected for the lens magnification and computed as average of the two independent estimates of the AGN bolometric luminosity, obtained through the 2-10 keV bolometric correction \citepads{2012MNRAS.425..623L} and from the continuum luminosity at 1450 $\SI{}{\angstrom}$ \citepads{2011ApJ...742...93A, 2012MNRAS.422..478R}.}}
\label{tab:energetics_UFOs}
\end{table*}%
\begin{figure}
\resizebox{\hsize}{!}{\includegraphics{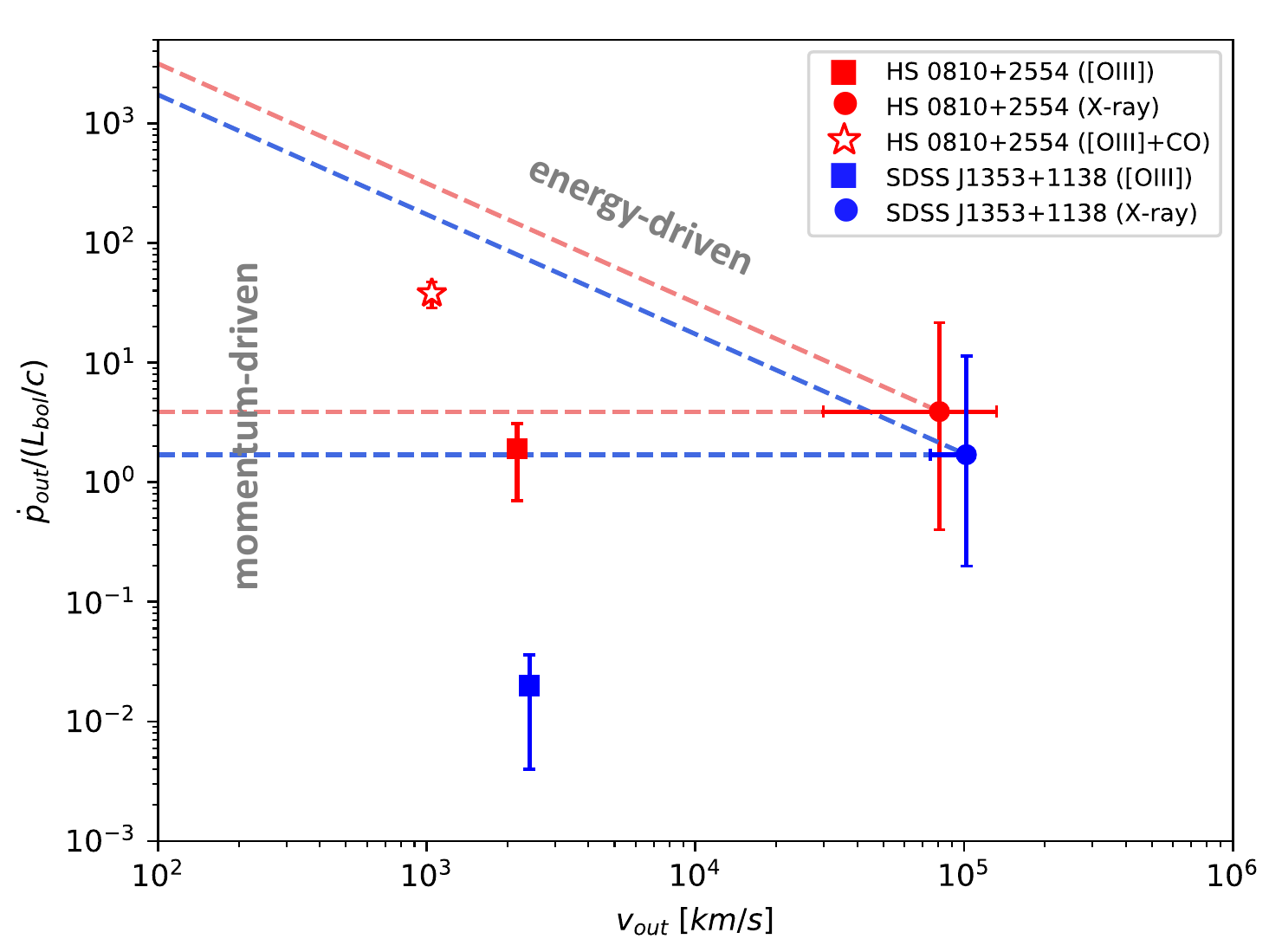}}
\caption{{\small Momentum-boost versus wind velocity diagram for both UFOs (represented as circles) and galactic outflow components (represented as squares) of HS 0810+2554 and SDSS J1353+1138. The dashed lines represent the predictions for a momentum-conserving regime (horizontal) and energy-conserving mode (diagonal). For HS 0810+2554, also the point relative to the ionised+molecular large-scale outflow is shown, identified by the star symbol.}}
\label{fig:pboost}
\end{figure}

Figure \ref{fig:pboost} shows the momentum-boost as a function of the outflow speed for HS 0810+2554 and SDSS J1353+1138. Given the two X-ray UFOs points (represented as circles), the predictions for a momentum-driven and an energy-driven scenario are represented by the dashed horizontal and diagonal lines, respectively. Errorbars on UFOs points indicate the minimum-maximum range of possible values, as  mentioned previously. In comparing with ionised outflow points (represented as squares), we observe that while the [O III] outflow of HS 0810+2554 is consistent (within the uncertainty) with the expectations for the momentum-driven mode, that of SDSS J1353+1138 is smaller by a factor of $\sim$100 than the predictions for a momentum-driven propagation, therefore, it is far from any tentative connection with the UFOs on nuclear scale. We reiterate that our measurement is a lower limit, as we did not detect [O III] emission from the second lensed image and, moreover, we approximated the outflow intrinsic radius to the one observed (just PSF-corrected). Nevertheless, even accounting for all these issues and approximations we made, such a discrepancy between observations and theoretical predictions could hardly be explained, as it amounts to about two orders of magnitude. Even associating half of the [O III] flux observed in image A to image B, on the basis of the flux ratio measurement of the two \textit{H}-band images \citepads{2006ApJ...653L..97I}, we would obtain a momentum-boost larger than the previous one by only a factor $\sim 1.5$. Similarly, it is unlikely that we are underestimating the stretching effects so much as to overestimate $R_{\rm out}$ by a factor of $\sim$100. Therefore, we conclude that such a significant discrepancy must have a different origin. The more plausible hypotheses are: 1) either the likely presence of a massive molecular outflow in this galaxy that our work is not accounting for; or 2) the possibility that the observed UFO is caused by an extraordinary burst episode (see e.g. \citeads{2020MNRAS.498.3633Z}) in the BH accretion activity of SDSS J1353+1138, while the large-scale outflow must be considered as the resultant effect of the AGN activity averaged over longer time-scales \citepads{2017ApJ...839..120W}. For this object, \citet{Chartas.high-z.UFOs} estimate a photon index $\Gamma \sim2.2$, which is typical of Narrow-line Seyfert 1 galaxies \citepads{1999ApJS..125..317L, 1999MNRAS.309..113V} and a typical signature of highly accreting systems \citepads{2019ApJ...876..102H}. The Eddington ratio $\lambda_{\rm Edd}$ (defined as $L_{\rm Bol}/L_{\rm Edd}$) estimated for SDSS J1353+1138 is $\lambda_{\rm Edd}=0.20\pm0.02$ \citep{Chartas.high-z.UFOs}, which is larger by a factor $\sim$3 than that inferred for HS 0810+2554 ($\lambda_{\rm Edd}=0.07\pm0.03$; \citealt{Chartas.high-z.UFOs}). Such results could support the recent post-burst scenario. Certainly, each hypothesis does not necessarily exclude the other and the real situation can be a combination of the two (effects).

The small values of momentum-boost ($\sim0.02-2$) and of kinetic efficiency ($\sim9-700\times10^{-5}$), inferred for the ionised outflows in SDSS J1353+1138 and HS 0810+2554, could be explained by an overall scarcity of [O III] (not only for the outflow component) in highly accreting AGNs, observed in other local galaxies with similar properties (e.g. \citeads{2000ApJ...536L...5S}). In our two QSOs, the poor outflow energetics is indeed due mainly to small values of outflow mass, and not to particularly low velocities. Such a scenario is related to the Eigenvector-1 effect: while the sources accreting at high rates (close to the Eddington limit) are actually the most promising candidates for hosting an active UFO (e.g. \citeads{2019MNRAS.482L.134N}), they usually present a very bright Fe II emission and a faint, outshined [O III] emission. As a consequence, the [O III] may be not ideal to trace the ionised phase of the outflow in AGNs accreting at high rates since the bulk of the ionised gas could be in the form of different chemical species.

It is also possible we underestimated the uncertainty of the [O III] outflow in HS 0810+2554, mostly due to our approximated procedure in the unlensed reconstruction. Moreover, given the recent tentative detection of a more massive CO-outflow on large scales claimed in \citetads{2020MNRAS.496..598C}, such a small value of the momentum-boost of the ionised outflow is not as unexpected given that it accounts only for the ionised gas traced by the [O III] emission. Hence, we also determined  the momentum-boost relative to the ionised plus molecular large-scale outflow ($p_{\rm out,~tot}/(L_{\rm Bol}/c)\sim38$), assuming its velocity to be equal to the mass-weighted average between the ionised and the CO-molecular outflow velocities ($v_{\rm out, tot}\sim1044$ km s$^{-1}$). Given the two orders of magnitude of difference between ionised and molecular outflow masses, the mass-weighted velocity is essentially the molecular outflow velocity ($v_{\rm out,~tot}=1040$ km s$^{-1}$; \citeads{2020MNRAS.496..598C}). 

In Fig. \ref{fig:pboost}, we report the CO+[O III] point with its uncertainty. Once  the contribution of the molecular component is included, the energetics of the overall large-scale outflow in HS 0810+2554 is compatible with an energy-driven scenario of wind propagation, within the (large) UFO uncertainty. However, deeper observations are required to confirm the CO-outflow detection and to constrain its energetics.

\subsection{Comparison with other QSOs hosting UFOs} \label{sec:comparison}

\begin{figure*}
\centering
\includegraphics[width=17cm,keepaspectratio]{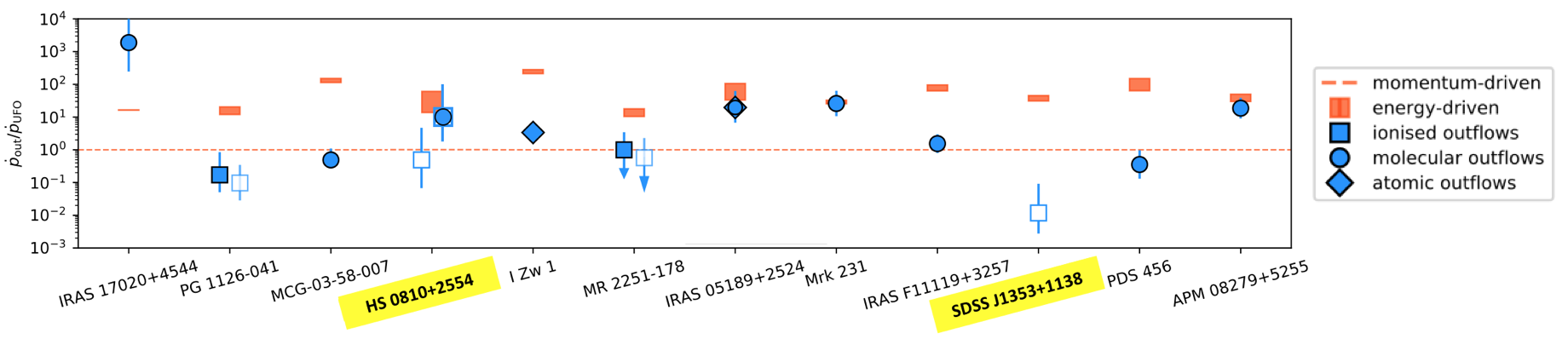}
\caption{{\small Ratio between the galaxy-scale and sub-pc scale outflow momentum rates for different QSOs hosting UFOs. Measurements for individual objects are shown in blue with the respective errorbars, using different markers according to the gas phase of the observed large scale outflow. The galaxy points are ordered by increasing $L_{\rm Bol}$. The horizontal dashed line shows the prediction for a momentum-driven wind ($\dot{p}_{\rm out}/\dot{p}_{\rm UFO}=1$), while the orange rectangles indicate the individual predictions for energy-driven winds. Filled and empty squares represent ionised outflow measurements based on H$\alpha$ and [O III] emission, respectively. For HS 0810+2554, our [O III]-based measurement is shown both alone and combined with the CO-measurement of the molecular outflow \citepads{2020MNRAS.496..598C}, with the respective symbol and combination of symbols (see the plot legend).}}
\label{fig:ratio}
\end{figure*}

Our study has revealed that the energetics of the galaxy-scale ionised outflow in HS 0810+2554 is consistent with the expectations of the momentum-driven mechanism, whereas in SDSS J1353+1138, the ionised outflow does not seem to be related with the detected UFO event on sub-pc scales. However, the exiguity of our two-QSOs sample prevents us from testing the predictions of theoretical models, as well as making general considerations on the nature of the mechanism powering outflows on large scales. Therefore, we considered our results along with those of a sample of well-studied QSOs, hosting both UFOs and galactic outflows, recently collected in \citetads{2020A&A...644A..15M}. The QSO-sample consists of the two local objects, MR225-178 and PG 1126-041, analysed in \citetads{2020A&A...644A..15M} to trace the ionised phase of the large-scale outflows, similarly to this work, but at low redshift; in addition, the other local QSOs having reliable UFOs and molecular or atomic outflow measurements. In terms of redshift, the only exception is the lensed QSO APM 08279+5255 at $z\sim3.9$ \citepads{2002ApJ...573L..77H}, which has been found to host a molecular outflow in energy-driven regime \citepads{2017A&A...608A..30F}.
In addition to HS 0810+2554 and SDSS J1353+1138, this is the only other case at high redshift, for which it has been possible to study the connection between the nuclear and the galaxy-scale winds. This highlights once again the importance of gravitational lensing as a powerful tool to overcome the limits imposed by the current observations.

\citetads{2020A&A...644A..15M} re-computed the UFO mass rate (and the wind energetics, consequently) for each QSO of their gathered sample, starting from the known estimates for $M_{\rm BH}$, $N_{\rm H}$, and $v_{\rm UFO}$. All measurements for molecular outflows have been re-scaled to the same luminosity-to-mass conversion factor $\alpha_{\rm CO}=0.8~(\text{K km s}^{-1}\text{pc}^2)^{-1}\text{ \(\text{M}_\odot\)}$, typical of QSOs, starburst and submillimeter galaxies \citepads{1998ApJ...507..615D, 2013ARA&A..51..207B, 2013ARA&A..51..105C}. The main AGN, X-ray-wind, and large-scale outflow properties of the known QSO-sample are listed in Table B.1 in \citetads{2020A&A...644A..15M}.

Figure \ref{fig:ratio} represents the updated version of Fig. 9 in \citetads{2020A&A...644A..15M}, including the measurements relative to HS 0810+2554 and SDSS J1353+1138: it shows the ratios between the galaxy-scale and sub-pc scale outflow momentum rates ($\dot{p}_{\rm out}$ and $\dot{p}_{\rm UFO}$, respectively), ordered according to the increasing $L_{\rm Bol}$ of the host AGN. This is an alternative way (with regard to Fig. \ref{fig:pboost}) to compare observational results with theoretical predictions for each QSO and to distinguish between the two main regimes. The energetics of the large-scale outflow in 10 out of 12 QSOs results to be consistent (or nearly consistent) with either a momentum-driven or an energy-driven regime:  in seven (six) and three (four) objects, respectively, if we exclude (include) the contribution of the tentatively detected molecular outflow in HS 0810+2554. This globally confirms the conclusion drawn by \citetads{2020A&A...644A..15M} that models of either momentum- or energy-driven outflows describe  the mechanism of wind propagation on galaxy-scales very well. The only two exceptions are SDSS J1353+1138 and IRAS 17020+4544, representative of the two extreme opposite situations in which the energetics on large and nuclear scales seem to be completely unrelated. As discussed in Sect. \ref{sec:connecting}, the low value of $\dot{p}_{\rm out}/\dot{p}_{\rm UFO}$ in SDSS J1353+1138 could be due to the presence of a massive molecular outflow, which our work does not account for. Alternatively, it could indicate that SDSS J1353+1138 has recently undergone a burst episode of its AGN activity. In terms of high AGN variability, the high value of $\dot{p}_{\rm out}/\dot{p}_{\rm UFO}$ inferred for IRAS 17020+4554 could be explained by invoking a past higher BH accretion activity compared to present day, but observations reveal that IRAS 17020+2554 is now accreting at a substantial fraction of its Eddington rate ($\lambda_{\rm Edd} \sim 0.7$; \citeads{2015ApJ...813L..39L}). Finally, we do not observe any remarkable trend in $\dot{p}_{\rm out}/\dot{p}_{\rm UFO}$ values with $L_{\rm Bol}$, nor any evident dependence of wind acceleration mechanism on galaxy redshift upon separately inspecting the results obtained for the high-redshift (our two objects plus APM 08279+5255) and low-redshift QSOs (the remaining ones) in the sample.

\section{Conclusions} \label{sec:conclusion}

Galaxy-wide outflows powered by AGN activity are thought to play a fundamental role in shaping the evolution of galaxies, as they allow us to reconcile theoretical models to observations. However, even though observations have widely confirmed their presence in both local and high-redshift galaxies, a clear understanding of the mechanism which accelerates these powerful, galaxy-scale winds is still lacking. To test the predictions of the current theoretical models, we need to compare, in a given object, the energetics of the sub-pc wind with that of the galaxy-wide outflow. The optimal sources in attempting the make a connection between different scales are the powerful QSOs near the peak of AGN activity ($z\sim2$), where the AGN feedback is expected to be more effective.

Given such a perspective, this work focuses on two $z\sim1.5$ multiple lensed QSOs, specifically selected to host UFOs (HS 0810+2554 and SDSS J1353+1138) and observed with the near-IR integral field spectrometer SINFONI. Thanks to the strong lens magnification and to the spatially resolved SINFONI data, which trace the dynamics of the ionised gas phase through rest-frame optical emission lines, it has become possible, for the first time, to attempt to make the connection between the sub-pc winds and the large-scale ionised outflows in two QSOs near the peak of AGN activity. The only other well-studied case at high redshift is APM 08279+5255 ($z\sim3.9$), which has been suggested to host a molecular outflow in energy-conserving regime \citepads{2017A&A...608A..30F}. Moreover, the recent analysis of ALMA data of HS 0810+2554 has revealed the tentative detection of CO-molecular outflow \citepads{2020MNRAS.496..598C}. Therefore, the characterisation of the ionised phase of the outflow in HS 0810+2554 provides the first three-phase description of an AGN-driven wind at high redshift, from the X-ray to the optical and mm bands, corresponding to highly ionised, ionised, and molecular gas phases, respectively.

In the following, we summarise the main results obtained in this work:
\begin{enumerate}
\item We studied the gas kinematics to identify the presence of outflowing gas by tracing the emission of the forbidden line doublet [O III]$\lambda \lambda$4959,5007, whose line profile is highly asymmetric in presence of outflows, with a typical broad, blueshifted wing corresponding to high speeds along the line of sight. In both QSO spectra, we detected the presence of extended ($\sim8$ kpc) ionised outflows moving up to $v\sim2000$ km s$^{-1}$ in the image lens plane.
\item After correcting for the gravitational lensing effects, we found that the ionised outflow in HS 0810+2554 is consistent within the uncertainty with the predictions for a momentum-driven regime and with an energy-driven propagation once  the contribution of the molecular outflow is included. On the contrary, the ionised outflow in SDSS J1353+1138 appears to be unrelated to the nuclear scale energetics, likely requiring either the presence of a massive molecular outflow or a high variability among the QSO activity.
\item By comparing our inferred results with those of the small sample of known QSOs from the literature, each hosting both sub-pc scale UFOs and neutral or ionised winds on galaxy scales, we found that the momentum- and energy-driven frameworks describe  all the observed targets very well, with the exception of SDSS J1353+1138 and IRAS 17020+4544. Therefore, these driving mechanisms appear to explain how the energy released by the AGN activity is coupled with the galactic ISM, thus driving the wind propagation on a large scale.
\end{enumerate}

Altogether, the observations presented in this work provide important pieces of information on the long sought-after ‘engine’ of large scale outflows and feedback, for the first time, at the crucial epoch for AGN feedback in galaxies, highlighting once again the power of integral field spectroscopy in this type of study. Follow-up CO observations of these sources will be necessary to confirm the molecular outflow detection and to constrain its energetics in HS 0810+2554 and to test whether a massive molecular outflow is responsible for the mismatch between the wind energetics at different scales observed in SDSS J1353+1138.

{\footnotesize \textit{Acknowledgments.} We thank the anonymous referee for comments and suggestions, which have improved the paper. We acknowledge support from the Italian Ministry for University and Research (MUR) for the BLACKOUT project funded through grant PRIN 2017PH3WAT.003. MP is supported by the Programa Atracción de Talento de la Comunidad de Madrid via grant 2018-T2/TIC-11715. MP acknowledges support from the Spanish Ministerio de Economía y Competitividad through the grant ESP2017-83197-P, and PID2019-106280GB-I00. Some data shown in this work were obtained from the Mikulski Archive for Space Telescopes (MAST).}

\bibliographystyle{aa} 
\bibliography{mybiblio} 

\begin{thebibliography}{122}
\expandafter\ifx\csname natexlab\endcsname\relax\def\natexlab#1{#1}\fi

\bibitem[{{Aird} {et~al.}(2015){Aird}, {Coil}, {Georgakakis}, {Nandra},
  {Barro}, \& {P{\'e}rez-Gonz{\'a}lez}}]{2015MNRAS.451.1892A}
{Aird}, J., {Coil}, A.~L., {Georgakakis}, A., {et~al.} 2015, \mnras, 451, 1892

\bibitem[{{Aird} {et~al.}(2010){Aird}, {Nandra}, {Laird}, {Georgakakis},
  {Ashby}, {Barmby}, {Coil}, {Huang}, {Koekemoer}, {Steidel}, \&
  {Willmer}}]{2010MNRAS.401.2531A}
{Aird}, J., {Nandra}, K., {Laird}, E.~S., {et~al.} 2010, \mnras, 401, 2531

\bibitem[{{Assef} {et~al.}(2011){Assef}, {Denney}, {Kochanek}, {Peterson},
  {Koz{\l}owski}, {Ageorges}, {Barrows}, {Buschkamp}, {Dietrich}, {Falco},
  {Feiz}, {Gemperlein}, {Germeroth}, {Grier}, {Hofmann}, {Juette}, {Khan},
  {Kilic}, {Knierim}, {Laun}, {Lederer}, {Lehmitz}, {Lenzen}, {Mall}, {Madsen},
  {Mandel}, {Martini}, {Mathur}, {Mogren}, {Mueller}, {Naranjo}, {Pasquali},
  {Polsterer}, {Pogge}, {Quirrenbach}, {Seifert}, {Stern}, {Shappee}, {Storz},
  {Van Saders}, {Weiser}, \& {Zhang}}]{2011ApJ...742...93A}
{Assef}, R.~J., {Denney}, K.~D., {Kochanek}, C.~S., {et~al.} 2011, \apj, 742,
  93

\bibitem[{{Bae} \& {Woo}(2018)}]{2018ApJ...853..185B}
{Bae}, H.-J. \& {Woo}, J.-H. 2018, \apj, 853, 185

\bibitem[{{Bautista} \& {Kallman}(2001)}]{2001ApJS..134..139B}
{Bautista}, M.~A. \& {Kallman}, T.~R. 2001, \apjs, 134, 139

\bibitem[{{Bayliss} {et~al.}(2017){Bayliss}, {Sharon}, {Acharyya}, {Gladders},
  {Rigby}, {Bian}, {Bordoloi}, {Runnoe}, {Dahle}, {Kewley}, {Florian},
  {Johnson}, \& {Paterno-Mahler}}]{2017ApJ...845L..14B}
{Bayliss}, M.~B., {Sharon}, K., {Acharyya}, A., {et~al.} 2017, \apjl, 845, L14

\bibitem[{{Bolatto} {et~al.}(2013){Bolatto}, {Wolfire}, \&
  {Leroy}}]{2013ARA&A..51..207B}
{Bolatto}, A.~D., {Wolfire}, M., \& {Leroy}, A.~K. 2013, \araa, 51, 207

\bibitem[{{Boroson} \& {Green}(1992)}]{1992ApJS...80..109B}
{Boroson}, T.~A. \& {Green}, R.~F. 1992, \apjs, 80, 109

\bibitem[{{Bouch{\'e}} {et~al.}(2013){Bouch{\'e}}, {Murphy}, {Kacprzak},
  {P{\'e}roux}, {Contini}, {Martin}, \&
  {Dessauges-Zavadsky}}]{2013Sci...341...50B}
{Bouch{\'e}}, N., {Murphy}, M.~T., {Kacprzak}, G.~G., {et~al.} 2013, Science,
  341, 50

\bibitem[{{Brusa} {et~al.}(2018){Brusa}, {Cresci}, {Daddi}, {Paladino},
  {Perna}, {Bongiorno}, {Lusso}, {Sargent}, {Casasola}, {Feruglio},
  {Fraternali}, {Georgiev}, {Mainieri}, {Carniani}, {Comastri}, {Duras},
  {Fiore}, {Mannucci}, {Marconi}, {Piconcelli}, {Zamorani}, {Gilli}, {La
  Franca}, {Lanzuisi}, {Lutz}, {Santini}, {Scoville}, {Vignali}, {Vito},
  {Rabien}, {Busoni}, \& {Bonaglia}}]{2018A&A...612A..29B}
{Brusa}, M., {Cresci}, G., {Daddi}, E., {et~al.} 2018, \aap, 612, A29

\bibitem[{{Brusa} {et~al.}(2016){Brusa}, {Perna}, {Cresci}, {Schramm},
  {Delvecchio}, {Lanzuisi}, {Mainieri}, {Mignoli}, {Zamorani}, {Berta},
  {Bongiorno}, {Comastri}, {Fiore}, {Kakkad}, {Marconi}, {Rosario}, {Contini},
  \& {Lamareille}}]{2016A&A...588A..58B}
{Brusa}, M., {Perna}, M., {Cresci}, G., {et~al.} 2016, \aap, 588, A58

\bibitem[{{Cano-D{\'\i}az} {et~al.}(2012){Cano-D{\'\i}az}, {Maiolino},
  {Marconi}, {Netzer}, {Shemmer}, \& {Cresci}}]{2012A&A...537L...8C}
{Cano-D{\'\i}az}, M., {Maiolino}, R., {Marconi}, A., {et~al.} 2012, \aap, 537,
  L8

\bibitem[{{Cappellari}(2017)}]{2017MNRAS.466..798C}
{Cappellari}, M. 2017, \mnras, 466, 798

\bibitem[{{Carilli} \& {Walter}(2013)}]{2013ARA&A..51..105C}
{Carilli}, C.~L. \& {Walter}, F. 2013, \araa, 51, 105

\bibitem[{{Carniani} {et~al.}(2015){Carniani}, {Marconi}, {Maiolino},
  {Balmaverde}, {Brusa}, {Cano-D{\'\i}az}, {Cicone}, {Comastri}, {Cresci},
  {Fiore}, {Feruglio}, {La Franca}, {Mainieri}, {Mannucci}, {Nagao}, {Netzer},
  {Piconcelli}, {Risaliti}, {Schneider}, \& {Shemmer}}]{2015A&A...580A.102C}
{Carniani}, S., {Marconi}, A., {Maiolino}, R., {et~al.} 2015, \aap, 580, A102

\bibitem[{{Carniani} {et~al.}(2016){Carniani}, {Marconi}, {Maiolino},
  {Balmaverde}, {Brusa}, {Cano-D{\'\i}az}, {Cicone}, {Comastri}, {Cresci},
  {Fiore}, {Feruglio}, {La Franca}, {Mainieri}, {Mannucci}, {Nagao}, {Netzer},
  {Piconcelli}, {Risaliti}, {Schneider}, \& {Shemmer}}]{2016A&A...591A..28C}
{Carniani}, S., {Marconi}, A., {Maiolino}, R., {et~al.} 2016, \aap, 591, A28

\bibitem[{{Carniani} {et~al.}(2017){Carniani}, {Marconi}, {Maiolino},
  {Feruglio}, {Brusa}, {Cresci}, {Cano-D{\'\i}az}, {Cicone}, {Balmaverde},
  {Fiore}, {Ferrara}, {Gallerani}, {La Franca}, {Mainieri}, {Mannucci},
  {Netzer}, {Piconcelli}, {Sani}, {Schneider}, {Shemmer}, \&
  {Testi}}]{2017A&A...605A.105C}
{Carniani}, S., {Marconi}, A., {Maiolino}, R., {et~al.} 2017, \aap, 605, A105

\bibitem[{{Cazzoli} {et~al.}(2016){Cazzoli}, {Arribas}, {Maiolino}, \&
  {Colina}}]{2016A&A...590A.125C}
{Cazzoli}, S., {Arribas}, S., {Maiolino}, R., \& {Colina}, L. 2016, \aap, 590,
  A125

\bibitem[{{Chartas} {et~al.}(2003){Chartas}, {Brandt}, \&
  {Gallagher}}]{2003ApJ...595...85C}
{Chartas}, G., {Brandt}, W.~N., \& {Gallagher}, S.~C. 2003, \apj, 595, 85

\bibitem[{{Chartas} {et~al.}(2002){Chartas}, {Brandt}, {Gallagher}, \&
  {Garmire}}]{2002ApJ...579..169C}
{Chartas}, G., {Brandt}, W.~N., {Gallagher}, S.~C., \& {Garmire}, G.~P. 2002,
  \apj, 579, 169

\bibitem[{{Chartas} {et~al.}(2016){Chartas}, {Cappi}, {Hamann}, {Eracleous},
  {Strickland}, {Giustini}, \& {Misawa}}]{2016ApJ...824...53C}
{Chartas}, G., {Cappi}, M., {Hamann}, F., {et~al.} 2016, \apj, 824, 53

\bibitem[{{Chartas} {et~al.}(2021){Chartas}, {Cappi}, {Vignali}, {Dadina},
  {James}, {Lanzuisi}, {Giustini}, {Gaspari}, {Strickland}, \&
  {Bertola}}]{Chartas.high-z.UFOs}
{Chartas}, G., {Cappi}, M., {Vignali}, C., {et~al.} 2021, \apj~(submitted)

\bibitem[{{Chartas} {et~al.}(2009){Chartas}, {Charlton}, {Eracleous},
  {Giustini}, {Hidalgo}, {Ganguly}, {Hamann}, {Misawa}, \&
  {Tytler}}]{2009NewAR..53..128C}
{Chartas}, G., {Charlton}, J., {Eracleous}, M., {et~al.} 2009, \nar, 53, 128

\bibitem[{{Chartas} {et~al.}(2020){Chartas}, {Davidson}, {Brusa}, {Vignali},
  {Cappi}, {Dadina}, {Cresci}, {Paladino}, {Lanzuisi}, \&
  {Comastri}}]{2020MNRAS.496..598C}
{Chartas}, G., {Davidson}, E., {Brusa}, M., {et~al.} 2020, \mnras, 496, 598

\bibitem[{{Chartas} {et~al.}(2007){Chartas}, {Eracleous}, {Dai}, {Agol}, \&
  {Gallagher}}]{2007ApJ...661..678C}
{Chartas}, G., {Eracleous}, M., {Dai}, X., {Agol}, E., \& {Gallagher}, S. 2007,
  \apj, 661, 678

\bibitem[{{Chartas} {et~al.}(2014){Chartas}, {Hamann}, {Eracleous}, {Misawa},
  {Cappi}, {Giustini}, {Charlton}, \& {Marvin}}]{2014ApJ...783...57C}
{Chartas}, G., {Hamann}, F., {Eracleous}, M., {et~al.} 2014, \apj, 783, 57

\bibitem[{{Cicone} {et~al.}(2018){Cicone}, {Brusa}, {Ramos Almeida}, {Cresci},
  {Husemann}, \& {Mainieri}}]{2018NatAs...2..176C}
{Cicone}, C., {Brusa}, M., {Ramos Almeida}, C., {et~al.} 2018, Nature
  Astronomy, 2, 176

\bibitem[{{Cicone} {et~al.}(2012){Cicone}, {Feruglio}, {Maiolino}, {Fiore},
  {Piconcelli}, {Menci}, {Aussel}, \& {Sturm}}]{2012A&A...543A..99C}
{Cicone}, C., {Feruglio}, C., {Maiolino}, R., {et~al.} 2012, \aap, 543, A99

\bibitem[{{Cicone} {et~al.}(2014){Cicone}, {Maiolino}, {Sturm},
  {Graci{\'a}-Carpio}, {Feruglio}, {Neri}, {Aalto}, {Davies}, {Fiore},
  {Fischer}, {Garc{\'\i}a-Burillo}, {Gonz{\'a}lez-Alfonso}, {Hailey-Dunsheath},
  {Piconcelli}, \& {Veilleux}}]{2014A&A...562A..21C}
{Cicone}, C., {Maiolino}, R., {Sturm}, E., {et~al.} 2014, \aap, 562, A21

\bibitem[{{Ciotti} {et~al.}(2010){Ciotti}, {Ostriker}, \&
  {Proga}}]{2010ApJ...717..708C}
{Ciotti}, L., {Ostriker}, J.~P., \& {Proga}, D. 2010, \apj, 717, 708

\bibitem[{{Coatman} {et~al.}(2017){Coatman}, {Hewett}, {Banerji}, {Richards},
  {Hennawi}, \& {Prochaska}}]{2017MNRAS.465.2120C}
{Coatman}, L., {Hewett}, P.~C., {Banerji}, M., {et~al.} 2017, \mnras, 465, 2120

\bibitem[{{Concas} {et~al.}(2017){Concas}, {Popesso}, {Brusa}, {Mainieri},
  {Erfanianfar}, \& {Morselli}}]{2017A&A...606A..36C}
{Concas}, A., {Popesso}, P., {Brusa}, M., {et~al.} 2017, \aap, 606, A36

\bibitem[{{Costa} {et~al.}(2014){Costa}, {Sijacki}, \&
  {Haehnelt}}]{2014MNRAS.444.2355C}
{Costa}, T., {Sijacki}, D., \& {Haehnelt}, M.~G. 2014, \mnras, 444, 2355

\bibitem[{{Cresci} {et~al.}(2009){Cresci}, {Hicks}, {Genzel}, {Schreiber},
  {Davies}, {Bouch{\'e}}, {Buschkamp}, {Genel}, {Shapiro}, {Tacconi},
  {Sommer-Larsen}, {Burkert}, {Eisenhauer}, {Gerhard}, {Lutz}, {Naab},
  {Sternberg}, {Cimatti}, {Daddi}, {Erb}, {Kurk}, {Lilly}, {Renzini},
  {Shapley}, {Steidel}, \& {Caputi}}]{2009ApJ...697..115C}
{Cresci}, G., {Hicks}, E.~K.~S., {Genzel}, R., {et~al.} 2009, \apj, 697, 115

\bibitem[{{Cresci} {et~al.}(2015){Cresci}, {Mainieri}, {Brusa}, {Marconi},
  {Perna}, {Mannucci}, {Piconcelli}, {Maiolino}, {Feruglio}, {Fiore},
  {Bongiorno}, {Lanzuisi}, {Merloni}, {Schramm}, {Silverman}, \&
  {Civano}}]{2015ApJ...799...82C}
{Cresci}, G., {Mainieri}, V., {Brusa}, M., {et~al.} 2015, \apj, 799, 82

\bibitem[{{Cresci} \& {Maiolino}(2018)}]{2018NatAs...2..179C}
{Cresci}, G. \& {Maiolino}, R. 2018, Nature Astronomy, 2, 179

\bibitem[{{Cresci} {et~al.}(2010){Cresci}, {Mannucci}, {Maiolino}, {Marconi},
  {Gnerucci}, \& {Magrini}}]{2010Natur.467..811C}
{Cresci}, G., {Mannucci}, F., {Maiolino}, R., {et~al.} 2010, \nat, 467, 811

\bibitem[{{Dadina} {et~al.}(2018){Dadina}, {Vignali}, {Cappi}, {Lanzuisi},
  {Ponti}, {Torresi}, {De Marco}, {Chartas}, \&
  {Giustini}}]{2018A&A...610L..13D}
{Dadina}, M., {Vignali}, C., {Cappi}, M., {et~al.} 2018, \aap, 610, L13

\bibitem[{{Davies} {et~al.}(2020){Davies}, {Baron}, {Shimizu}, {Netzer},
  {Burtscher}, {de Zeeuw}, {Genzel}, {Hicks}, {Koss}, {Lin}, {Lutz},
  {Maciejewski}, {M{\"u}ller-S{\'a}nchez}, {Orban de Xivry}, {Ricci}, {Riffel},
  {Riffel}, {Rosario}, {Schartmann}, {Schnorr-M{\"u}ller}, {Shangguan},
  {Sternberg}, {Sturm}, {Storchi-Bergmann}, {Tacconi}, \&
  {Veilleux}}]{2020MNRAS.498.4150D}
{Davies}, R., {Baron}, D., {Shimizu}, T., {et~al.} 2020, \mnras, 498, 4150

\bibitem[{{Di Matteo} {et~al.}(2005){Di Matteo}, {Springel}, \&
  {Hernquist}}]{2005Natur.433..604D}
{Di Matteo}, T., {Springel}, V., \& {Hernquist}, L. 2005, \nat, 433, 604

\bibitem[{{Downes} \& {Solomon}(1998)}]{1998ApJ...507..615D}
{Downes}, D. \& {Solomon}, P.~M. 1998, \apj, 507, 615

\bibitem[{{Eisenhauer} {et~al.}(2003){Eisenhauer}, {Abuter}, {Bickert},
  {Biancat-Marchet}, {Bonnet}, {Brynnel}, {Conzelmann}, {Delabre}, {Donaldson},
  {Farinato}, {Fedrigo}, {Genzel}, {Hubin}, {Iserlohe}, {Kasper},
  {Kissler-Patig}, {Monnet}, {Roehrle}, {Schreiber}, {Stroebele}, {Tecza},
  {Thatte}, \& {Weisz}}]{2003SPIE.4841.1548E}
{Eisenhauer}, F., {Abuter}, R., {Bickert}, K., {et~al.} 2003, in Society of
  Photo-Optical Instrumentation Engineers (SPIE) Conference Series, Vol. 4841,
  Instrument Design and Performance for Optical/Infrared Ground-based
  Telescopes, ed. M.~{Iye} \& A.~F.~M. {Moorwood}, 1548--1561

\bibitem[{{Faber} \& {Jackson}(1976)}]{1976ApJ...204..668F}
{Faber}, S.~M. \& {Jackson}, R.~E. 1976, \apj, 204, 668

\bibitem[{{Fabian}(2012)}]{2012ARA&A..50..455F}
{Fabian}, A.~C. 2012, \araa, 50, 455

\bibitem[{{Ferland} {et~al.}(2009){Ferland}, {Hu}, {Wang}, {Baldwin}, {Porter},
  {van Hoof}, \& {Williams}}]{2009ApJ...707L..82F}
{Ferland}, G.~J., {Hu}, C., {Wang}, J.-M., {et~al.} 2009, \apjl, 707, L82

\bibitem[{{Ferrarese} \& {Merritt}(2000)}]{2000ApJ...539L...9F}
{Ferrarese}, L. \& {Merritt}, D. 2000, \apjl, 539, L9

\bibitem[{{Feruglio} {et~al.}(2017){Feruglio}, {Ferrara}, {Bischetti},
  {Downes}, {Neri}, {Ceccarelli}, {Cicone}, {Fiore}, {Gallerani}, {Maiolino},
  {Menci}, {Piconcelli}, {Vietri}, {Vignali}, \&
  {Zappacosta}}]{2017A&A...608A..30F}
{Feruglio}, C., {Ferrara}, A., {Bischetti}, M., {et~al.} 2017, \aap, 608, A30

\bibitem[{{Feruglio} {et~al.}(2015){Feruglio}, {Fiore}, {Carniani},
  {Piconcelli}, {Zappacosta}, {Bongiorno}, {Cicone}, {Maiolino}, {Marconi},
  {Menci}, {Puccetti}, \& {Veilleux}}]{2015A&A...583A..99F}
{Feruglio}, C., {Fiore}, F., {Carniani}, S., {et~al.} 2015, \aap, 583, A99

\bibitem[{{Feruglio} {et~al.}(2010){Feruglio}, {Maiolino}, {Piconcelli},
  {Menci}, {Aussel}, {Lamastra}, \& {Fiore}}]{2010A&A...518L.155F}
{Feruglio}, C., {Maiolino}, R., {Piconcelli}, E., {et~al.} 2010, \aap, 518,
  L155

\bibitem[{{Fiore} {et~al.}(2017){Fiore}, {Feruglio}, {Shankar}, {Bischetti},
  {Bongiorno}, {Brusa}, {Carniani}, {Cicone}, {Duras}, {Lamastra}, {Mainieri},
  {Marconi}, {Menci}, {Maiolino}, {Piconcelli}, {Vietri}, \&
  {Zappacosta}}]{2017A&A...601A.143F}
{Fiore}, F., {Feruglio}, C., {Shankar}, F., {et~al.} 2017, \aap, 601, A143

\bibitem[{{Fluetsch} {et~al.}(2020){Fluetsch}, {Maiolino}, {Carniani},
  {Arribas}, {Belfiore}, {Bellocchi}, {Cazzoli}, {Cicone}, {Cresci}, {Fabian},
  {Gallagher}, {Ishibashi}, {Mannucci}, {Marconi}, {Perna}, {Sturm}, \&
  {Venturi}}]{2020arXiv200613232F}
{Fluetsch}, A., {Maiolino}, R., {Carniani}, S., {et~al.} 2020, arXiv e-prints,
  arXiv:2006.13232

\bibitem[{{Fluetsch} {et~al.}(2019){Fluetsch}, {Maiolino}, {Carniani},
  {Marconi}, {Cicone}, {Bourne}, {Costa}, {Fabian}, {Ishibashi}, \&
  {Venturi}}]{2019MNRAS.483.4586F}
{Fluetsch}, A., {Maiolino}, R., {Carniani}, S., {et~al.} 2019, \mnras, 483,
  4586

\bibitem[{{F{\"o}rster Schreiber} {et~al.}(2019){F{\"o}rster Schreiber},
  {{\"U}bler}, {Davies}, {Genzel}, {Wisnioski}, {Belli}, {Shimizu}, {Lutz},
  {Fossati}, {Herrera-Camus}, {Mendel}, {Tacconi}, {Wilman}, {Beifiori},
  {Brammer}, {Burkert}, {Carollo}, {Davies}, {Eisenhauer}, {Fabricius},
  {Lilly}, {Momcheva}, {Naab}, {Nelson}, {Price}, {Renzini}, {Saglia},
  {Sternberg}, {van Dokkum}, \& {Wuyts}}]{2019ApJ...875...21F}
{F{\"o}rster Schreiber}, N.~M., {{\"U}bler}, H., {Davies}, R.~L., {et~al.}
  2019, \apj, 875, 21

\bibitem[{Fruchter \& Hook(2019)}]{drizzle_book_bis}
Fruchter, A.~S. \& Hook, R.~N. 2019, Drizzle: A Method for the Linear
  Reconstruction of Undersampled Images (PASP), 114--144

\bibitem[{{Gofford} {et~al.}(2015){Gofford}, {Reeves}, {McLaughlin}, {Braito},
  {Turner}, {Tombesi}, \& {Cappi}}]{2015MNRAS.451.4169G}
{Gofford}, J., {Reeves}, J.~N., {McLaughlin}, D.~E., {et~al.} 2015, \mnras,
  451, 4169

\bibitem[{{Gofford} {et~al.}(2013){Gofford}, {Reeves}, {Tombesi}, {Braito},
  {Turner}, {Miller}, \& {Cappi}}]{2013MNRAS.430...60G}
{Gofford}, J., {Reeves}, J.~N., {Tombesi}, F., {et~al.} 2013, \mnras, 430, 60

\bibitem[{Gonzaga {et~al.}(2012)Gonzaga, Hack, Fruchter, \&
  Mack}]{drizzle_book}
Gonzaga, S., Hack, W., Fruchter, A., \& Mack, J. 2012, The DrizzlePac Handbook
  (Baltimore, STScI)

\bibitem[{{Granato} {et~al.}(2004){Granato}, {De Zotti}, {Silva}, {Bressan}, \&
  {Danese}}]{2004ApJ...600..580G}
{Granato}, G.~L., {De Zotti}, G., {Silva}, L., {Bressan}, A., \& {Danese}, L.
  2004, \apj, 600, 580

\bibitem[{{Guo} {et~al.}(2016){Guo}, {Gonzalez-Perez}, {Guo}, {Schaller},
  {Furlong}, {Bower}, {Cole}, {Crain}, {Frenk}, {Helly}, {Lacey}, {Lagos},
  {Mitchell}, {Schaye}, \& {Theuns}}]{2016MNRAS.461.3457G}
{Guo}, Q., {Gonzalez-Perez}, V., {Guo}, Q., {et~al.} 2016, \mnras, 461, 3457

\bibitem[{{Harrison} {et~al.}(2014){Harrison}, {Alexander}, {Mullaney}, \&
  {Swinbank}}]{2014MNRAS.441.3306H}
{Harrison}, C.~M., {Alexander}, D.~M., {Mullaney}, J.~R., \& {Swinbank}, A.~M.
  2014, \mnras, 441, 3306

\bibitem[{{Harrison} {et~al.}(2018){Harrison}, {Costa}, {Tadhunter},
  {Fl{\"u}tsch}, {Kakkad}, {Perna}, \& {Vietri}}]{2018NatAs...2..198H}
{Harrison}, C.~M., {Costa}, T., {Tadhunter}, C.~N., {et~al.} 2018, Nature
  Astronomy, 2, 198

\bibitem[{{Hasinger} {et~al.}(2002){Hasinger}, {Schartel}, \&
  {Komossa}}]{2002ApJ...573L..77H}
{Hasinger}, G., {Schartel}, N., \& {Komossa}, S. 2002, \apjl, 573, L77

\bibitem[{{Huang} {et~al.}(2019){Huang}, {Hu}, {Zhao}, {Zhang}, {Lu}, {Wang},
  {Zhang}, {Du}, {Li}, {Bai}, {Ho}, {Bian}, {Yuan}, \&
  {Wang}}]{2019ApJ...876..102H}
{Huang}, Y.-K., {Hu}, C., {Zhao}, Y.-L., {et~al.} 2019, \apj, 876, 102

\bibitem[{{Inada} {et~al.}(2006){Inada}, {Oguri}, {Morokuma}, {Doi}, {Yasuda},
  {Becker}, {Richards}, {Kochanek}, {Kayo}, {Konishi}, {Utsunomiya}, {Shin},
  {Strauss}, {Sheldon}, {York}, {Hennawi}, {Schneider}, {Dai}, \&
  {Fukugita}}]{2006ApJ...653L..97I}
{Inada}, N., {Oguri}, M., {Morokuma}, T., {et~al.} 2006, \apjl, 653, L97

\bibitem[{{Jackson} {et~al.}(2015){Jackson}, {Tagore}, {Roberts}, {Sluse},
  {Stacey}, {Vives-Arias}, {Wucknitz}, \& {Volino}}]{2015MNRAS.454..287J}
{Jackson}, N., {Tagore}, A.~S., {Roberts}, C., {et~al.} 2015, \mnras, 454, 287

\bibitem[{{Kakkad} {et~al.}(2020){Kakkad}, {Mainieri}, {Vietri}, {Carniani},
  {Harrison}, {Perna}, {Scholtz}, {Circosta}, {Cresci}, {Husemann},
  {Bischetti}, {Feruglio}, {Fiore}, {Marconi}, {Padovani}, {Brusa}, {Cicone},
  {Comastri}, {Lanzuisi}, {Mannucci}, {Menci}, {Netzer}, {Piconcelli},
  {Puglisi}, {Salvato}, {Schramm}, {Silverman}, {Vignali}, {Zamorani}, \&
  {Zappacosta}}]{2020A&A...642A.147K}
{Kakkad}, D., {Mainieri}, V., {Vietri}, G., {et~al.} 2020, \aap, 642, A147

\bibitem[{{Kaspi} {et~al.}(2000){Kaspi}, {Smith}, {Netzer}, {Maoz}, {Jannuzi},
  \& {Giveon}}]{2000ApJ...533..631K}
{Kaspi}, S., {Smith}, P.~S., {Netzer}, H., {et~al.} 2000, \apj, 533, 631

\bibitem[{{King}(2003)}]{2003ApJ...596L..27K}
{King}, A. 2003, \apjl, 596, L27

\bibitem[{{King}(2005)}]{2005ApJ...635L.121K}
{King}, A. 2005, \apjl, 635, L121

\bibitem[{{King} \& {Pounds}(2015)}]{2015ARA&A..53..115K}
{King}, A. \& {Pounds}, K. 2015, \araa, 53, 115

\bibitem[{{King}(2010{\natexlab{a}})}]{2010MNRAS.408L..95K}
{King}, A.~R. 2010{\natexlab{a}}, \mnras, 408, L95

\bibitem[{{King}(2010{\natexlab{b}})}]{2010MNRAS.402.1516K}
{King}, A.~R. 2010{\natexlab{b}}, \mnras, 402, 1516

\bibitem[{{Kormendy} \& {Ho}(2013)}]{2013ARA&A..51..511K}
{Kormendy}, J. \& {Ho}, L.~C. 2013, \araa, 51, 511

\bibitem[{{Kova{\v{c}}evi{\'c}} {et~al.}(2010){Kova{\v{c}}evi{\'c}},
  {Popovi{\'c}}, \& {Dimitrijevi{\'c}}}]{2010ApJS..189...15K}
{Kova{\v{c}}evi{\'c}}, J., {Popovi{\'c}}, L.~{\v{C}}., \& {Dimitrijevi{\'c}},
  M.~S. 2010, \apjs, 189, 15

\bibitem[{{Lanzuisi} {et~al.}(2012){Lanzuisi}, {Giustini}, {Cappi}, {Dadina},
  {Malaguti}, {Vignali}, \& {Chartas}}]{2012A&A...544A...2L}
{Lanzuisi}, G., {Giustini}, M., {Cappi}, M., {et~al.} 2012, \aap, 544, A2

\bibitem[{{Law} {et~al.}(2009){Law}, {Steidel}, {Erb}, {Larkin}, {Pettini},
  {Shapley}, \& {Wright}}]{2009ApJ...697.2057L}
{Law}, D.~R., {Steidel}, C.~C., {Erb}, D.~K., {et~al.} 2009, \apj, 697, 2057

\bibitem[{{Leighly}(1999)}]{1999ApJS..125..317L}
{Leighly}, K.~M. 1999, \apjs, 125, 317

\bibitem[{{Longinotti} {et~al.}(2015){Longinotti}, {Krongold}, {Guainazzi},
  {Giroletti}, {Panessa}, {Costantini}, {Santos-Lleo}, \&
  {Rodriguez-Pascual}}]{2015ApJ...813L..39L}
{Longinotti}, A.~L., {Krongold}, Y., {Guainazzi}, M., {et~al.} 2015, \apjl,
  813, L39

\bibitem[{{Ludwig} {et~al.}(2009){Ludwig}, {Wills}, {Greene}, \&
  {Robinson}}]{2009ApJ...706..995L}
{Ludwig}, R.~R., {Wills}, B., {Greene}, J.~E., \& {Robinson}, E.~L. 2009, \apj,
  706, 995

\bibitem[{{Lusso} {et~al.}(2012){Lusso}, {Comastri}, {Simmons}, {Mignoli},
  {Zamorani}, {Vignali}, {Brusa}, {Shankar}, {Lutz}, {Trump}, {Maiolino},
  {Gilli}, {Bolzonella}, {Puccetti}, {Salvato}, {Impey}, {Civano}, {Elvis},
  {Mainieri}, {Silverman}, {Koekemoer}, {Bongiorno}, {Merloni}, {Berta}, {Le
  Floc'h}, {Magnelli}, {Pozzi}, \& {Riguccini}}]{2012MNRAS.425..623L}
{Lusso}, E., {Comastri}, A., {Simmons}, B.~D., {et~al.} 2012, \mnras, 425, 623

\bibitem[{{Madau} \& {Dickinson}(2014)}]{2014ARA&A..52..415M}
{Madau}, P. \& {Dickinson}, M. 2014, \araa, 52, 415

\bibitem[{{Madau} {et~al.}(1996){Madau}, {Ferguson}, {Dickinson}, {Giavalisco},
  {Steidel}, \& {Fruchter}}]{1996MNRAS.283.1388M}
{Madau}, P., {Ferguson}, H.~C., {Dickinson}, M.~E., {et~al.} 1996, \mnras, 283,
  1388

\bibitem[{{Maiolino} {et~al.}(2012){Maiolino}, {Gallerani}, {Neri}, {Cicone},
  {Ferrara}, {Genzel}, {Lutz}, {Sturm}, {Tacconi}, {Walter}, {Feruglio},
  {Fiore}, \& {Piconcelli}}]{2012MNRAS.425L..66M}
{Maiolino}, R., {Gallerani}, S., {Neri}, R., {et~al.} 2012, \mnras, 425, L66

\bibitem[{{Marasco} {et~al.}(2020){Marasco}, {Cresci}, {Nardini}, {Mannucci},
  {Marconi}, {Tozzi}, {Tozzi}, {Amiri}, {Venturi}, {Piconcelli}, {Lanzuisi},
  {Tombesi}, {Mingozzi}, {Perna}, {Carniani}, {Brusa}, \& {di Serego
  Alighieri}}]{2020A&A...644A..15M}
{Marasco}, A., {Cresci}, G., {Nardini}, E., {et~al.} 2020, \aap, 644, A15

\bibitem[{{Marconi} {et~al.}(2004){Marconi}, {Risaliti}, {Gilli}, {Hunt},
  {Maiolino}, \& {Salvati}}]{2004MNRAS.351..169M}
{Marconi}, A., {Risaliti}, G., {Gilli}, R., {et~al.} 2004, \mnras, 351, 169

\bibitem[{{Mosquera} \& {Kochanek}(2011)}]{2011ApJ...738...96M}
{Mosquera}, A.~M. \& {Kochanek}, C.~S. 2011, \apj, 738, 96

\bibitem[{{Nagao} {et~al.}(2006){Nagao}, {Marconi}, \&
  {Maiolino}}]{2006A&A...447..157N}
{Nagao}, T., {Marconi}, A., \& {Maiolino}, R. 2006, \aap, 447, 157

\bibitem[{{Narayan} \& {Bartelmann}(1996)}]{1996astro.ph..6001N}
{Narayan}, R. \& {Bartelmann}, M. 1996, arXiv e-prints, astro

\bibitem[{{Nardini} {et~al.}(2019){Nardini}, {Lusso}, \&
  {Bisogni}}]{2019MNRAS.482L.134N}
{Nardini}, E., {Lusso}, E., \& {Bisogni}, S. 2019, \mnras, 482, L134

\bibitem[{{Nardini} {et~al.}(2015){Nardini}, {Reeves}, {Gofford}, {Harrison},
  {Risaliti}, {Braito}, {Costa}, {Matzeu}, {Walton}, {Behar}, {Boggs},
  {Christensen}, {Craig}, {Hailey}, {Matt}, {Miller}, {O'Brien}, {Stern},
  {Turner}, \& {Ward}}]{2015Sci...347..860N}
{Nardini}, E., {Reeves}, J.~N., {Gofford}, J., {et~al.} 2015, Science, 347, 860

\bibitem[{{Nardini} \& {Zubovas}(2018)}]{2018MNRAS.478.2274N}
{Nardini}, E. \& {Zubovas}, K. 2018, \mnras, 478, 2274

\bibitem[{{Nelson} {et~al.}(2019){Nelson}, {Pillepich}, {Springel}, {Pakmor},
  {Weinberger}, {Genel}, {Torrey}, {Vogelsberger}, {Marinacci}, \&
  {Hernquist}}]{2019MNRAS.490.3234N}
{Nelson}, D., {Pillepich}, A., {Springel}, V., {et~al.} 2019, \mnras, 490, 3234

\bibitem[{{Nierenberg} {et~al.}(2020){Nierenberg}, {Gilman}, {Treu}, {Brammer},
  {Birrer}, {Moustakas}, {Agnello}, {Anguita}, {Fassnacht}, {Motta}, {Peter},
  \& {Sluse}}]{2020MNRAS.492.5314N}
{Nierenberg}, A.~M., {Gilman}, D., {Treu}, T., {et~al.} 2020, \mnras, 492, 5314

\bibitem[{{Osterbrock}(1981)}]{1981ApJ...249..462O}
{Osterbrock}, D.~E. 1981, \apj, 249, 462

\bibitem[{Oya {et~al.}(2016)Oya, Terada, Hayano, Watanabe, Hattori, \&
  Minowa}]{cite-key}
Oya, S., Terada, H., Hayano, Y., {et~al.} 2016, Experimental Astronomy, 42, 85

\bibitem[{{Peng} {et~al.}(2006){Peng}, {Impey}, {Rix}, {Kochanek}, {Keeton},
  {Falco}, {Leh{\'a}r}, \& {McLeod}}]{2006ApJ...649..616P}
{Peng}, C.~Y., {Impey}, C.~D., {Rix}, H.-W., {et~al.} 2006, \apj, 649, 616

\bibitem[{{Perna} {et~al.}(2015){Perna}, {Brusa}, {Cresci}, {Comastri},
  {Lanzuisi}, {Lusso}, {Marconi}, {Salvato}, {Zamorani}, {Bongiorno},
  {Mainieri}, {Maiolino}, \& {Mignoli}}]{2015A&A...574A..82P}
{Perna}, M., {Brusa}, M., {Cresci}, G., {et~al.} 2015, \aap, 574, A82

\bibitem[{{Perna} {et~al.}(2017){Perna}, {Lanzuisi}, {Brusa}, {Cresci}, \&
  {Mignoli}}]{2017A&A...606A..96P}
{Perna}, M., {Lanzuisi}, G., {Brusa}, M., {Cresci}, G., \& {Mignoli}, M. 2017,
  \aap, 606, A96

\bibitem[{{Pontzen} {et~al.}(2017){Pontzen}, {Tremmel}, {Roth}, {Peiris},
  {Saintonge}, {Volonteri}, {Quinn}, \& {Governato}}]{2017MNRAS.465..547P}
{Pontzen}, A., {Tremmel}, M., {Roth}, N., {et~al.} 2017, \mnras, 465, 547

\bibitem[{{Reimers} {et~al.}(2002){Reimers}, {Hagen}, {Baade}, {Lopez}, \&
  {Tytler}}]{2002A&A...382L..26R}
{Reimers}, D., {Hagen}, H.~J., {Baade}, R., {Lopez}, S., \& {Tytler}, D. 2002,
  \aap, 382, L26

\bibitem[{{Rizzo} {et~al.}(2018){Rizzo}, {Vegetti}, {Fraternali}, \& {Di
  Teodoro}}]{2018MNRAS.481.5606R}
{Rizzo}, F., {Vegetti}, S., {Fraternali}, F., \& {Di Teodoro}, E. 2018, \mnras,
  481, 5606

\bibitem[{{Ross} {et~al.}(2009){Ross}, {Assef}, {Kochanek}, {Falco}, \&
  {Poindexter}}]{2009ApJ...702..472R}
{Ross}, N.~R., {Assef}, R.~J., {Kochanek}, C.~S., {Falco}, E., \& {Poindexter},
  S.~D. 2009, \apj, 702, 472

\bibitem[{{Runnoe} {et~al.}(2012){Runnoe}, {Brotherton}, \&
  {Shang}}]{2012MNRAS.422..478R}
{Runnoe}, J.~C., {Brotherton}, M.~S., \& {Shang}, Z. 2012, \mnras, 422, 478

\bibitem[{{Rupke} {et~al.}(2005){Rupke}, {Veilleux}, \&
  {Sanders}}]{2005ApJS..160..115R}
{Rupke}, D.~S., {Veilleux}, S., \& {Sanders}, D.~B. 2005, \apjs, 160, 115

\bibitem[{{Rusu} {et~al.}(2016){Rusu}, {Oguri}, {Minowa}, {Iye}, {Inada},
  {Oya}, {Kayo}, {Hayano}, {Hattori}, {Saito}, {Ito}, {Pyo}, {Terada},
  {Takami}, \& {Watanabe}}]{2016MNRAS.458....2R}
{Rusu}, C.~E., {Oguri}, M., {Minowa}, Y., {et~al.} 2016, \mnras, 458, 2

\bibitem[{{Silk} \& {Rees}(1998)}]{1998A&A...331L...1S}
{Silk}, J. \& {Rees}, M.~J. 1998, \aap, 331, L1

\bibitem[{{Spingola} {et~al.}(2020){Spingola}, {McKean}, {Vegetti}, {Powell},
  {Auger}, {Koopmans}, {Fassnacht}, {Lagattuta}, {Rizzo}, {Stacey}, \&
  {Sweijen}}]{2020MNRAS.495.2387S}
{Spingola}, C., {McKean}, J.~P., {Vegetti}, S., {et~al.} 2020, \mnras, 495,
  2387

\bibitem[{{Stacey} {et~al.}(2021){Stacey}, {McKean}, {Powell}, {Vegetti},
  {Rizzo}, {Spingola}, {Auger}, {Ivison}, \& {van der
  Werf}}]{2021MNRAS.500.3667S}
{Stacey}, H.~R., {McKean}, J.~P., {Powell}, D.~M., {et~al.} 2021, \mnras, 500,
  3667

\bibitem[{{Steidel} {et~al.}(2010){Steidel}, {Erb}, {Shapley}, {Pettini},
  {Reddy}, {Bogosavljevi{\'c}}, {Rudie}, \& {Rakic}}]{2010ApJ...717..289S}
{Steidel}, C.~C., {Erb}, D.~K., {Shapley}, A.~E., {et~al.} 2010, \apj, 717, 289

\bibitem[{{Sulentic} {et~al.}(2000){Sulentic}, {Zwitter}, {Marziani}, \&
  {Dultzin-Hacyan}}]{2000ApJ...536L...5S}
{Sulentic}, J.~W., {Zwitter}, T., {Marziani}, P., \& {Dultzin-Hacyan}, D. 2000,
  \apjl, 536, L5

\bibitem[{{Tombesi} {et~al.}(2012){Tombesi}, {Cappi}, {Reeves}, \&
  {Braito}}]{2012MNRAS.422L...1T}
{Tombesi}, F., {Cappi}, M., {Reeves}, J.~N., \& {Braito}, V. 2012, \mnras, 422,
  L1

\bibitem[{{Tombesi} {et~al.}(2013){Tombesi}, {Cappi}, {Reeves}, {Nemmen},
  {Braito}, {Gaspari}, \& {Reynolds}}]{2013MNRAS.430.1102T}
{Tombesi}, F., {Cappi}, M., {Reeves}, J.~N., {et~al.} 2013, \mnras, 430, 1102

\bibitem[{{Tombesi} {et~al.}(2011){Tombesi}, {Cappi}, {Reeves}, {Palumbo},
  {Braito}, \& {Dadina}}]{2011ApJ...742...44T}
{Tombesi}, F., {Cappi}, M., {Reeves}, J.~N., {et~al.} 2011, \apj, 742, 44

\bibitem[{{Tombesi} {et~al.}(2010){Tombesi}, {Cappi}, {Reeves}, {Palumbo},
  {Yaqoob}, {Braito}, \& {Dadina}}]{2010A&A...521A..57T}
{Tombesi}, F., {Cappi}, M., {Reeves}, J.~N., {et~al.} 2010, \aap, 521, A57

\bibitem[{{Vaughan} {et~al.}(1999){Vaughan}, {Reeves}, {Warwick}, \&
  {Edelson}}]{1999MNRAS.309..113V}
{Vaughan}, S., {Reeves}, J., {Warwick}, R., \& {Edelson}, R. 1999, \mnras, 309,
  113

\bibitem[{{Vegetti} \& {Koopmans}(2009)}]{2009MNRAS.392..945V}
{Vegetti}, S. \& {Koopmans}, L.~V.~E. 2009, \mnras, 392, 945

\bibitem[{{Vignali} {et~al.}(2015){Vignali}, {Iwasawa}, {Comastri}, {Gilli},
  {Lanzuisi}, {Ranalli}, {Cappelluti}, {Mainieri}, {Georgantopoulos},
  {Carrera}, {Fritz}, {Brusa}, {Brandt}, {Bauer}, {Fiore}, \&
  {Tombesi}}]{2015A&A...583A.141V}
{Vignali}, C., {Iwasawa}, K., {Comastri}, A., {et~al.} 2015, \aap, 583, A141

\bibitem[{{Woo} {et~al.}(2017){Woo}, {Son}, \& {Bae}}]{2017ApJ...839..120W}
{Woo}, J.-H., {Son}, D., \& {Bae}, H.-J. 2017, \apj, 839, 120

\bibitem[{{Zakamska} \& {Greene}(2014)}]{2014MNRAS.442..784Z}
{Zakamska}, N.~L. \& {Greene}, J.~E. 2014, \mnras, 442, 784

\bibitem[{{Zubovas} \& {King}(2012)}]{2012ApJ...745L..34Z}
{Zubovas}, K. \& {King}, A. 2012, \apjl, 745, L34

\bibitem[{{Zubovas} \& {King}(2014)}]{2014MNRAS.439..400Z}
{Zubovas}, K. \& {King}, A.~R. 2014, \mnras, 439, 400

\bibitem[{{Zubovas} \& {Nardini}(2020)}]{2020MNRAS.498.3633Z}
{Zubovas}, K. \& {Nardini}, E. 2020, \mnras, 498, 3633

\end{thebibliography}

\begin{appendix}
\section{Approximated reconstruction of the unlensed outflow in HS 0810+2554} \label{app:2D_rec}

As mentioned in Sect. \ref{sec:intrinsic}, for HS 0810+2554 we were able to obtain a partial 2D-reconstruction of the outflow emission in the source plane by using the gravitational lens fitting-code presented in \citetads{2018MNRAS.481.5606R}, which adopts the lensing operator described in \citetads{2009MNRAS.392..945V}. This depends, in particular, on lensing operators describing the lens as a power law plus a shear component (see Sect. 2.2 in \citeads{2018MNRAS.481.5606R} for details).

We began by simulating possible intrinsic geometries of the outflow. We considered multiple conical configurations, differing in radius and aperture angle, while the position angle of the cone was fixed to $\theta \sim130^{\circ}$, namely, the direction in which we observed the major [O III] outflow emission in the lens plane (see Sect. \ref{sec:test_spatially_res}). In order to reduce further the number of free parameters characterising the outflow geometry, we fixed the origin of all simulated cones to the position of the emission centroid of the unlensed outflow, obtained through a first tentative full 2D-reconstruction with the lens fitting-code by \citetads{2018MNRAS.481.5606R} and a lens model fixed to that found in \citetads{2020MNRAS.492.5314N}. In fact, even though it could not be used to establish the intrinsic extent of the [O III] outflow because of the presence of residual effects from the PSF-deconvolution, it still provided  reliable constraints of the centre position of the background outflow emission.
Each cone was modelled as a uniform distribution of homogeneous point-like sources, and then each point-source was individually lensed forward (i.e. singularly mapped and magnified in the lens plane) through the reconstruction-algorithm of \citetads{2018MNRAS.481.5606R}, keeping the lens model fixed to \citetads{2020MNRAS.492.5314N}. As a result, we obtained the whole forward lensed image of each starting background cone, differing in radius and aperture angle.

\begin{figure}
\resizebox{\hsize}{!}{\includegraphics{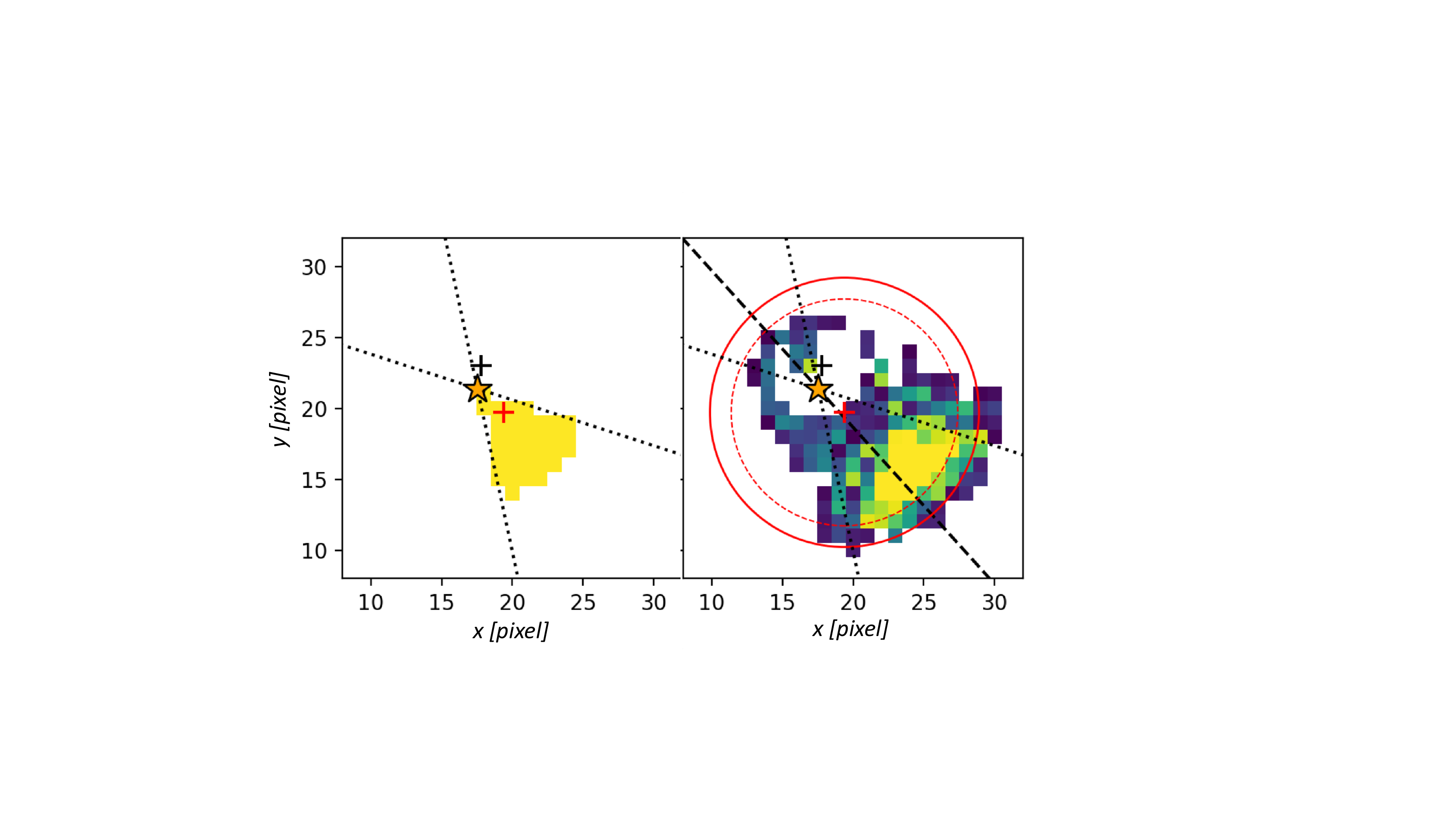}}  
\caption{{\small Example of forward lensing \citepads{2018MNRAS.481.5606R} of a background homogeneous cone with radius of 8 pixels and aperture of 60$^{\circ}$ (left) into the respective lensed image (right). In both images, the orange stars, red and black ‘+’ indicate, respectively, the position of the cone origin, of the emission peak observed by SINFONI, and of the lens centre. The dashed and solid red circumferences in the right panel have a radius equal to the intrinsic size of the cone in the source plane (i.e. 8 pixel in the example shown here) and to the maximum extent reached by the [O III] emission observed by SINFONI, i.e. $R_{\rm max}=9.5$ pixels; both circumferences are centred in the observed emission peak (red ‘+’). The dashed and dotted black lines identify the position angle ($\sim130^{\circ}$) and the aperture (here $60^{\circ}$) of the intrinsic cone. We note that the conical source does not intercept the centre of the lens, thus subject to a low magnification.}}
\label{fig:cone}
\end{figure}

In order to establish the intrinsic size of the outflow, we selected those cones with a radius compatible with the detected [O III] outflow emission by visually comparing the extent of the forward lensed emission produced by a given radius background cone, with the maximum distance at which we observe the [O III] outflow, once corrected for the SINFONI-PSF, that is, $R_{\rm max} = 9.5$ pixels\footnote{For simplicity, as it is the estimate of the outflow radius based on a qualitative comparison, we report the projected distances in the description
of the procedure in units of SINFONI pixels, thus recalling that $R_{\rm max}\sim8.7$ kpc corresponds to $\sim8$ SINFONI pixels. At the end, we will provide the estimate of the intrinsic size of the outflow in physical units.}. In Fig. \ref{fig:cone}, we show  an example of our forward lensing method: starting, for instance, from a background homogeneous cone with radius of 8 pixels and aperture of 60$^{\circ}$ (left) and its forward lensed image (right). The solid red circumference has a radius equal to $R_{\rm max}=9.5$ pixels, thus indicating the maximal extent reached by the observed (PSF-corrected) [O III] outflow emission (see caption of Fig. \ref{fig:cone} for a detailed description). In this way, we found that the cones with a radius ranging from 6.5 to 9.5 pixels were compatible with $R_{\rm max}$. Taking the average of these more plausible radii and the maximum deviation from the mean as error, and converting into kpc-units, we estimated the intrinsic radius of the outflow to be $R_{\rm out}=(8.7\pm1.7)$ kpc, with $z=1.508\pm0.002$ as the redshift we measured from the nuclear spectrum extracted during the BLR-fitting (described in Sect. \ref{sec:BLR_fit}).

For the estimate of the magnification factor with its error, we considered multiple aperture angles: 30$^{\circ}$, 45$^{\circ}$, 60$^{\circ}$, and 90$^{\circ}$, for each defined-radius cone, and calculated the magnification factor as the ratio between the total flux in the lens plane and the total flux in the source plane. All simulated conical configurations provided low magnification factors that are weakly dependent on the assumed geometry of the cone (they differ by a factor $\sim$1.4 at most). This follows from the fact that no simulated cones intercept the lens caustics and extend to large scales, where the lens magnification is reasonably lower. Therefore, to determine the outflow total mean magnification, we focused only on the magnification values obtained in correspondence to the established range of more plausible outflow radii, that is, $6.5-9.5$ pixels, and averaged over them, finding $\mu_{\rm out}=2.0\pm0.2$, where the maximum deviation from the mean has been taken as error. We used this value to correct the observed [O III] outflow flux (determined in Sect. \ref{sec:test_spatially_res}), thus obtaining $F_{\rm out}=(1.9\pm0.2)\times10^{-15}(2.0/\mu_{\rm out})$ erg s$^{-1}$cm$^{-2}$.

\end{appendix}
\listofobjects
\end{document}